\documentclass[twocolumn]{aa}
\usepackage{graphicx}
\usepackage{natbib}

\usepackage{txfonts}
\begin{document}
   \title{X-rays from T Tau: A test case for accreting T Tauri stars}

   \author{Manuel G\"udel
          \inst{1}
	  \and
	  Stephen L. Skinner
	  \inst{2}
	  \and
	  Stanislav Yu. Mel'nikov
	  \inst{3}
	  \and
	  Marc Audard
	  \inst{4}
          \fnmsep\thanks{\emph{New address (since September
          2006): }Integral Science Data Centre, Ch. d'Ecogia 16, CH-1290 Versoix, 
          Switzerland \& Geneva Observatory,  University of Geneva, Ch. des 
          Maillettes 51, 1290 Sauverny,    Switzerland}
	  \and
	  Alessandra Telleschi
	  \inst{1}
	  \and
	  Kevin R. Briggs
	  \inst{1}
          }

   \offprints{Manuel G\"udel}

   \institute{Paul Scherrer Institut, W\"urenlingen and Villigen,
              CH-5232 Villigen PSI, Switzerland\\
              \email{guedel@astro.phys.ethz.ch}
	\and 
	    CASA, 389,
	     University of Colorado,
	     Boulder, CO 80309-0389,
	     USA   
	\and
	      Ulugh Beg Astronomical Institute, 
	      Academy of Sciences of Uzbekistan, 
	      Astronomicheskaya ul. 33, 
	      Tashkent, 700052, Uzbekistan
	\and
	     Columbia Astrophysics Laboratory,
	      Mail Code 5247, 550 West 120th Street,
	      New York, NY 10027,
	      USA     
             }

   \date{Received 2006; accepted 2006}

  \abstract
   {The generation of X-rays in accreting T Tauri stars (TTS) is thought to be predominantly due to 
   energy dissipation in magnetic fields, but alternative X-ray generation mechanisms have been proposed,
   such as heating in accretion shocks near the stellar surface, or in shocks forming in jets.}
   {We test models and trends discussed in the  literature using X-ray data from the 
   classical TTS T Tau.}
   {High-resolution spectroscopy from the Reflection Grating Spectrometers on {{\it XMM-Newton}}
   is used to infer electron densities, element abundances and the thermal structure of the
   X-ray source. We also
   discuss the ultraviolet light curve obtained by the Optical Monitor, and complementary ground-based
   photometry. A high-resolution image
   from {{\it Chandra}} constrains contributions from the two companions of T Tau N.}
   {The X-ray grating spectrum is rich in emission lines, but shows an unusual mixture of features
   from very hot ($\approx 30$~MK) and very cool (1--3~MK) plasma, both emitted by similar amounts of
   emission measure. The cool plasma confirms the picture of a {{\it soft excess}} in
   the form of an enhanced O\,{\sc vii}/O\,{\sc viii} Ly$\alpha$ flux ratio, similar to that
   previously reported for other accreting TTS. Diagnostics from lines formed by this plasma 
   indicate low electron densities ($\la 10^{10}$~cm$^{-3}$). The Ne/Fe abundance ratio is consistent 
   with a trend in pre-main sequence
   stars in which this ratio depends on spectral type, but not on accretion.
   }
   {On the basis of line density diagnostics, we conclude that
    the density of the cool ``soft-excess'' plasma is orders of magnitude
    below that predicted for an accretion shock, assuming previously determined
    accretion rates of $(3-6)\times 10^{-8}M_{\odot}$yr$^{-1}$. We argue that loading of
    magnetic field lines with infalling material suppresses  
    the heating process in a part of the corona. We thus suggest that the X-ray production of T Tau 
    is influenced by the accretion process although the X-rays may not form in the bulk of the 
    accretion footpoints.}
   \keywords{Stars: coronae --
	     Stars: formation --
	     Stars: pre-main sequence --
	     X-rays: stars  --
	     Stars: individual: T Tau
	       }

   \maketitle
%
%________________________________________________________________

\section{Introduction}\label{introduction}

Low-mass pre-main sequence stars  maintain magnetic fields that are thought
to provide the  energy for high-energy particles and radiation as shown
by their strong X-ray and radio emission (for reviews, see, e.g., \citealt{feigelson99,
feigelson06, guedel02, guedel04}). Vigorous sources of X-ray radiation have been
found among class I protostars (e.g., \citealt{koyama94, imanishi01}), accreting classical
T Tau stars (CTTS, e.g., \citealt{walter81, feigelson81}), and weak-line T Tau
stars (WTTS; e.g., \citealt{walter86, neuhaeuser95}) that are no longer surrounded 
by a thick, accreting disk. A number of observational features have supported
a picture in which X-ray emission is due to magnetic coronal processes similar
to mechanisms seen in the solar corona:  rotational modulation 
points to a concentration of hot plasma in various magnetic active regions \citep{flaccomio05};
flares  heat plasma on time scales of minutes to hours, reaching temperatures
of a few tens of MK \citep{wolk05}; and electron temperatures continuously exceed $10$~MK (e.g., 
\citealt{imanishi01, skinner03}); such temperatures cannot be achieved by liberating gravitational 
energy from infall, nor by shocks in moving gas (e.g., jets) commonly observed 
around young stars at other wavelengths. Also, high-energy electrons observed through their gyrosynchrotron
emission give direct evidence for particle acceleration, for which magnetic
reconnection processes provide an ideal basis.

Nevertheless, accreting pre-main sequence stars add circumstellar disks and accretion
flows to the picture of the stellar environment; stellar magnetic fields
may reach out to the inner border of the disk from where they guide lightly ionized
mass to the stellar surface \citep{calvet98}. WTTS and CTTS
are therefore ideal objects to test to what extent the stellar environment controls
coronal magnetic fields and energy release. Except for disks and mass accretion streams, the two types of 
stars appear to be identical in nature, occupying the same region in the 
Hertzsprung-Russell Diagram (HRD).

Evidence for {\it some sort} of influence of the disk environment onto X-ray production 
is broad: i) Star-disk magnetic fields may brake stellar rotation, thus damping the
magnetic dynamo and inducing weaker overall magnetic activity. In the Taurus Molecular
Cloud, for example, CTTS rotate on average less rapidly than WTTS (characteristic
rotation periods of $P \approx 8$~d and 2--4~d, respectively; \citealt{bouvier93}). 
Many T Tau stars may, however, emit at the empirical saturation limit, $L_{\rm X}/L_{\rm bol} 
\approx 10^{-3.5}$,  for  rotation periods below 10~d \citep{preibisch05}, so that 
a difference between the dynamo operation of CTTS and WTTS may be difficult to identify. ii) Despite 
possible saturation effects, it has now been clearly established that both the X-ray 
luminosity, $L_{\rm X}$, and $L_{\rm X}/L_{\rm bol}$ are, for otherwise similar stellar characteristics,
{\it lower} by a factor of $\approx 2$ for CTTS than for WTTS 
\citep{neuhaeuser95, preibisch05, telleschi06a}. It is unclear in what sense accretion is
responsible for this result. iii) On the other hand, the electron temperature
of the X-ray emitting plasma may be  higher in CTTS than in WTTS \citep{imanishi01, tsujimoto02, telleschi06a}.
iv) Specific changes in the time behavior of X-rays were also observed in outburst phenomena
ascribed to strong accretion events in FU Ori-type and EX Lupi-type variables (e.g., a strong
flux increase of hard emission during outburst, \citealt{kastner04a, kastner06}; or a softening of
the plasma during outburst, \citealt{audard05}).

Recently, a more direct role of accretion has been suggested for the generation
of X-rays in CTTS, namely emission from gas that has been shock-heated in the
accretion flow near the stellar surface. While free-fall velocities of a few 100~km~s$^{-1}$ 
on CTTS are sufficient to heat shocks to a few MK, a  problem is X-ray absorption by the gas 
stream itself, and by the surrounding photosphere if the shock is buried below the surface 
\citep{calvet98, lamzin99, drake05}. However, the temperatures attained in such shocks
are clearly not sufficient to explain the extremely hot X-ray sources of CTTS  with temperatures 
exceeding 10~MK.
On the other hand, in the  exceptional case of the CTTS TW Hya, X-ray spectroscopic
evidence for a predominantly cool plasma led \citet{kastner02} to suggest 
that X-ray production by accretion may be relevant in this star. This conjecture has
been further supported by evidence for unusually high electron densities of $n_{\rm e} >
10^{11}$~cm$^{-3}$ derived from  density-sensitive line flux ratios of O\,{\sc vii} and  Ne\,{\sc ix} 
in high-resolution X-ray spectra of the CTTS TW Hya, BP Tau, and V4046 Sgr
 \citep{kastner02, stelzer04, schmitt05, robrade06, gunther06}, supported by flux ratios
of Fe\,{\sc xvii} lines \citep{ness05}. \citet{stelzer04} further
argued that high Ne/Fe and N/Fe abundance ratios may point to accreting gas that is
depleted of Fe, presumably because Fe (and several other elements) easily condense onto 
dust grains further out in the disk, leaving an accreting gas phase enriched by Ne and N. 

Two large survey projects of X-ray emission in star-forming regions have addressed the
above questions anew. \citet{preibisch05} found no support for an accretion scenario in 
in the {\it Chandra Orion Ultradeep Project} (COUP); the bulk of the 
observed X-ray emission is related to very hot, coronal plasma. Accretion may in fact suppress 
the X-ray luminosity by mass loading of magnetic fields. High-resolution X-ray spectroscopy
collected in the framework of the {\it XMM-Newton Extended Survey of the Taurus Molecular Cloud}
(XEST) has shown a series of new facets \citep{telleschi06b}: 
i) very high ($\ga 10^{11}$~cm$^{-3}$) electron densities are not recorded in all 
accreting pre-main sequence stars, posing problems for the accretion scenario even in the most
favorable case of   100\% accretion surface filling. ii) Abundance ``anomalies'', in particular
a high Ne/Fe abundance ratio,  are common to many T Tau stars, regardless of whether or not
they are accreting. On the other hand, the Ne/Fe abundance ratio seems to depend on the the stellar
spectral type. iii) While the bulk plasma observed in CTTS shows very similar thermal
properties as plasma in WTTS, there is {\it soft excess} emission in accreting stars; this 
excess is visible as (and is defined by) an enhanced ratio of the O\,{\sc vii}/O\,{\sc viii} 
Ly$\alpha$ line fluxes  compared with WTTS or zero-age main-sequence (ZAMS) stars. Because 
the maximum formation temperatures of O\,{\sc vii} and O\,{\sc viii} are 2~MK and 3~MK, respectively 
(and O\,{\sc viii} is formed up to considerably higher temperatures),
an enhanced ratio indicates the presence of additional plasma below 3~MK.
Observation  of this excess requires high-resolution spectroscopy that resolves emission 
lines formed at 1--4~MK such as lines from  O\,{\sc vii} or N\,{\sc vi} and N\,{\sc vii}.

To further our understanding of these issues, detailed case studies  
are needed. Ideally, a bright source will produce  detailed 
high-resolution grating spectra of the line-dominated range below 1~keV. 
This allows us to assess anomalies in the thermal structure and electron densities from line ratios.
To access lines formed at low (1--4~MK) temperatures (located predominantly at wavelengths
longward of 17~\AA), the photoelectric absorption should be moderate (hydrogen column density 
$N_{\rm H}$ of no more  than a few times
$10^{21}$~cm$^{-2}$). Only few CTTS fulfill these requirements. One of the most outstanding 
among them is the CTTS T Tau N. We report detailed X-ray spectroscopic
and imaging studies of this object, also searching for emission related to its two embedded
companions, T Tau Sab. 

The specific goal of this paper is to use T Tau as a test case for hypotheses of
X-ray production related to magnetic and  accretion processes. We add complementary information
from optical and ultraviolet observations. The plan of the paper is as follows.
In Sect.~\ref{target}, we introduce the T Tau system, and describe our observations
in Sect.~\ref{observations}. Sect.~\ref{image}
presents our imaging studies, Sect.~\ref{light} introduces light curves, and Sect.~\ref{spectroscopy}
presents results from our spectroscopic analysis. Sect.~\ref{discussion} contains a 
discussion and our conclusions.

\section{Target}\label{target}

Historically, the optically revealed northern component, T Tau N, was the prototype 
of accreting low-mass pre-main sequence stars.   Subsequently, a strongly variable IR companion was
found $0\farcs 7$ to its south,  now designated as T Tau S.
It was soon recognized to be a protostar probably co-eval with 
T Tau N \citep{dyck82} and dominating the system flux at $\lambda > 2\mu$m. 
The southern component was  later  found to be double, with a separation of 
only $0\farcs 09$ or 13~AU \citep{koresko00, duchene02}. An orbital analysis of
astrometric radio and infrared data infers masses of $(2.73\pm 0.31)~M_{\odot}$ 
and $(0.61\pm 0.17)~M_{\odot}$ for Tau Sa and Sb, respectively \citep{duchene06}, 
suggesting that the former is an embedded Herbig Ae star. The mass of T Tau N is  also higher than
the masses of typical T Tau stars, namely $M \approx 2.4~M_{\odot}$ \citep{guedel06a}, 
and its rotation period is $P = 2.8$~d \citep{herbst86}.  The distance to T Tau is $\approx$140~pc
\citep{loinard05}.

The T Tau triple may be at a critical transition stage between protostars and
fully optically revealed T Tau stars; some of its observational properties appear to be 
related to its particular orientation that places us into an optimal position
for observations. \citet{vanlangevelde94a} and \citet{vanlangevelde94b}  found evidence 
for infall of molecular  material and a collimated outflow,  pointing at 
13$^{\circ}$ to the observer. \citet{solf99} suggest a pole-on view
for T Tau N, which is also clearly supported 
by mm observations of a well-developed, near-face-on  massive accretion disk  
\citep{akeson98}. The latter authors point out that the disk is similar in mass 
to the solar nebula in which our planetary system  formed. The pole-on view  
of the N component results in a relatively low extinction, while from larger angles
all T Tau components would appear to be embedded accreting young stars. A  CO 
emission ring, identified with the outer walls of a bi-conical outflow region, provides 
evidence that we are looking down into a low-extinction cavity which the stellar 
wind has nearly swept free of molecular gas \citep{momose96}. 
T Tau Sa appears to be surrounded by an opaque disk seen nearly edge-on 
\citep{duchene05, duchene06}. The disk then is no more than 3~AU in radius given the
close orbit of the Sb companion (with a semi-major axis of 11.6~AU and a periastron distance
of $\approx 6$~AU).

The T Tau system has been detected as a  radio source 
with properties of  an extended wind but also of non-thermal radio emission 
\citep{phillips93, skinner94}, including gyrosynchrotron emission from an 
outflow (Ray et al. 1997) and strongly polarized emission from magnetic fields inferred to be
of kGauss strength \citep{smith03, loinard05}.
The T Tau system was well detected in  X-rays with the {\it ROSAT} PSPC 
 \citep{neuhaeuser95, stelzer00}, with a
quiescent count rate of (0.03$\pm$0.009)~cts~s$^{-1}$, and multiple energetic and rapid 
flares developing on  $\sim$1~hour timescales and reaching up to 0.2~cts~s$^{-1}$. 

\section{Observations and data reduction}\label{observations}

Table~\ref{tab:obslog} summarizes basic parameters of our {\it XMM-Newton} and {\it Chandra}
X-ray observations, including the centroid coordinates of the X-ray images of T Tau. Both exposures
comprise data of four stellar/substellar systems that are thought to be members of the Taurus
Molecular Cloud: these are, apart from the T Tau system, the M0 CTTS IRAS 04187+1927, the WTTS
RX~J0422.1+1934, and the brown dwarf 2M~J04221332+1934. All have been detected as X-ray 
sources. Further details on the former two stars are summarized in \cite{guedel06a},
while the brown dwarf is described in \citet{grosso06}.

\begin{table}[t!]
\centering
\caption{X-ray observing log and coordinates of T Tau}
\label{tab:obslog}
\begin{tabular}{lllll}
\hline
\noalign{\smallskip}
            Parameter                  &     {\it XMM-Newton}           &   {\it Chandra}                          \\                
            \noalign{\smallskip}
            \hline
            \noalign{\smallskip}
	    XEST number             &  01-045  		              & ....				\\		     
	    ObsID                   &  0301500101		      &  3732				     \\
	    Start time  (UT)        &  2005-08-15\ 13:52:13	      & 2003-10-24\ 06:11:52			\\
	    End time (UT)           &  2005-08-16\ 12:55:22	      & 2003-10-24\ 13:37:02			\\
	    Exposure time           &  82989~s			      & 26710~s  				  \\
	    Coordinates:$^a$        &                                 &                               \\
	    Boresight RA            &  04h 21m 59.4s                  &  04h 21m 59.69s                          \\
	    Boresight decl.         &  $19\deg\ 32\arcmin\ 06\arcsec$ &  $19\deg\ 32\arcmin\ 20\farcs 6$         \\
	    X-ray: RA               &  04h  21m  59.44s               &  04h 21m 59.42s	                  \\
	    X-ray: decl.            & $19\deg\ 32\arcmin\ 05\farcs 8$ & $19\deg\ 32\arcmin\ 06\farcs 6$       \\
	    T Tau N: RA$^{b}$       &  04h 21m 59.43s                 &  04h 21m 59.43s	                  \\
	    T Tau N: decl.$^{b}$    &  $19\deg\ 32\arcmin\ 06\farcs 4$& $19\deg\ 32\arcmin\ 06\farcs 4$       \\
            \hline
\end{tabular}
\begin{list}{}{}
\item[$^{\mathrm{a}}$] All coordinates are J2000.0
\item[$^{\mathrm{b}}$] For origin of coordinates, see \citet{smith03}
\end{list}
\end{table}

We obtained an $\approx $83~ks exposure of the T Tau system on
August 15--16, 2005, with {\it XMM-Newton} \citep{jansen01} using standard observing 
setups.\footnote{see {\it XMM-Newton} User's Handbook (http://xmm.vilspa.esa.es/)}
Specifically, the European Photon Imaging Cameras (EPIC) 
obtained images with an angular resolution of $\approx 4-5\arcsec$ (FWHM), using
the full window with a diameter of 30\arcmin\ for the PN camera \citep{strueder01}, while
the MOS1 and MOS2 cameras \citep{turner01} observed in the large-window (W3) and small-window (W2)  
configurations that constrain the central CCD to a field of 5.5\arcmin$\times$5.5\arcmin and  
1.8\arcmin$\times$1.8\arcmin, respectively. These constrained window modes were
chosen to prevent the detectors from pile-up effects in case of strong flares (but no strong
flares that would induce pile-up were actually observed). All EPIC cameras
used the medium blocking filter.

The two Reflection Grating Spectrometers (RGS; \citealt{denherder01}) were operating
in spectroscopy mode, providing  X-ray spectra with a resolution
of approximately 60--70~m\AA\ (FWHM) in the wavelength range of 5--35~\AA. Two chip failures
earlier in the mission left a gap in the  RGS1 spectrum in the range of 
10.6-13.8~\AA\, and in the RGS2 spectrum in the range of 20.0--24.1~\AA.

The Optical Monitor (OM; \citealt{mason01}) observed in imaging and high-time resolution 
mode with the UVW1 ultraviolet filter inserted. The latter transmits light roughly in the
2500--3500~\AA\ range. The imaging field of view is a square with a size of 17\arcmin$\times$17\arcmin.
The high-time resolution recording applies, however, only to a central window of 
10.5\arcsec$\times$10.5\arcsec\  in which T Tau was positioned.

The X-ray data analysis of the {\it XMM-Newton} observations is described in detail
by \citet{guedel06a}. For details on the OM data reduction, see \citet{audard06}. 
The data reduction made use of standard SAS v.6.1.0 software (i.e., the epchain, emchain, rgsproc, 
omfchain, and omichain tasks). Spectra and images were extracted with the evselect task.
For the EPIC data, we used canned response matrices appropriate for SAS v.6.1.0, whereas
ancillary response files (ARFs) were generated using the SAS task arfgen. The RGS
response matrix was generated with the rgsrmfgen task. Background radiation  fluctuated between 
a high and a low level during the observation, each occupying approximately
half of the time. The high particle radiation can affect accurate background subtraction 
in both RGS, in particular in long-wavelength regions where the observed, true continuum
is  low and suppression of the source spectrum by photoelectric absorption
may be severe. We have experimented with either extracting only low-background data, or with
constraining the width of the source extraction region in the RGS detector plane. We found that
extracting 70\% (down from standard 90\%) of the PSF width was optimum, and this choice 
was  preferred to selecting in time while keeping the extraction width
at the standard value.

The nearby WTTS RX~J0422.1+1934 was bright enough (amounting to  41\% of the count rate of T Tau in the EPIC detectors,  
see \citealt{guedel06a}) that it slightly contaminates the background region on the 
RGS detector. The source region of this star was therefore excluded  from the background region
of T Tau. It does not directly contaminate the T Tau spectrum, however, as it is
located at the edge of the RGS detector, far outside the standard spectral extraction width for T Tau.

We further obtained a {\it Chandra X-Ray Observatory} \citep{weisskopf02} exposure of 26~ks duration, using
the High Resolution Camera (HRC-I, \citealt{murray00}) with a spatial resolution of $\approx 0.5\arcsec$ but no
energy resolution in the X-ray range. The HRC field of view is about 30\arcmin\ in diameter.
The data were reduced using standard CIAO v.3.0.2a software 
tools\footnote{http://cxc.harvard.edu/ciao/guides/hrc\_data.html}.
New level 1 events were created with the hrc\_process\_events task, and level 2 events
after applying filters and good time interval (GTI) selection. A light curve
analysis revealed no significant variability.  A formal source detection was performed
with the wavdetect algorithm in order to obtain precise centroid coordinates for
the T Tau source. 
 
We obtained ground-based optical observations of T Tau during three nights 
around the {\it Chandra} observations. 
 Photometric monitoring  was performed in the UBVR bands with the 48-cm AZT-14 
reflector at the Mt. Maidanak Observatory during the nights of  
2003 October 22/23, 23/24, and 24/25. For these observations we used a single-channel 
photoelectric photometer in pulse-counting mode, and the photometric data were 
reduced to Johnson's international system. We observed HD~27560 as a 
comparison star and HD 284415 as check star. Other observational 
information can be found in \citet{melnikov05}.

\section{A high-resolution X-ray image of T Tau}\label{image}

The T Tau triple with a separation of 0\farcs 7 between the  N and Sab components cannot
be resolved by the {\it XMM-Newton} EPIC cameras. The {\it Chandra} HRC image, on the
other hand, could marginally reveal binary structure if the N and S components showed
comparable flux levels.  The centroid coordinate of the stellar X-ray image
(Table~\ref{tab:obslog}) is very close to T Tau N.

\begin{figure}[t!]
%\centering
\includegraphics[angle=0,width=7.8cm]{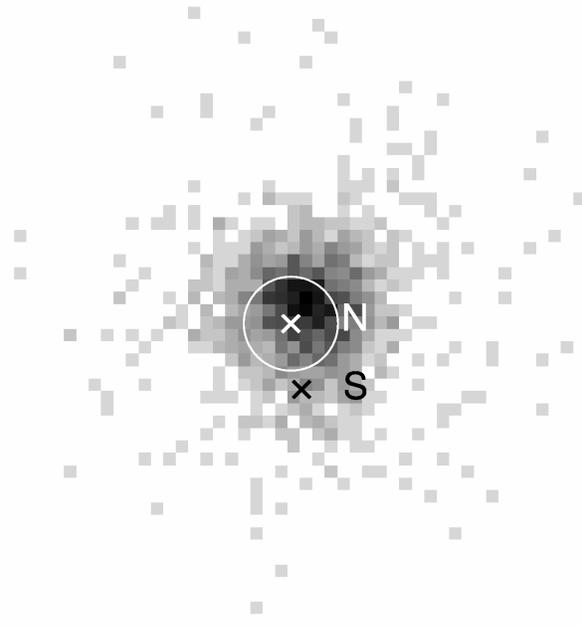}
%\vskip -1truecm
\caption{The T Tau source on the {\it Chandra} HRC-I detector (pixel size: $0\farcs 13$).
The positions of T Tau N and Sab are marked by the crosses. The white circle around the expected
T Tau N position indicates the 90\% source location error for {\it Chandra}. There is a slight systematic
offset of about $0\farcs 27$ of the X-ray image relative to the true position of  T Tau N
($-0\farcs 11$ in RA, $+0\farcs 25$  in declination in the sense ``X-ray minus true''), 
which has not been corrected for in this figure.}
\label{ttauimage} 
\end{figure}

Fig.~\ref{ttauimage} shows the HRC-I image of the T Tau source. We  modeled
a point-spread function (PSF) using the CIAO task mkpsf. For this, we used a circle
with a radius of 15 pixels (with a pixel size of 0.13\arcsec) around the centroid. 
The PSF was then
fitted to T Tau image in the {\it Chandra} sherpa software \citep{freeman01} 
(the version implemented in CIAO version 3.3.0.1 was used). The model also contained 
a spatially constant background.  We found the centroid to be slightly
but insignificantly offset from the true stellar position  
(offset   ``X-ray minus true'' of $-0\farcs 11$ in RA and $0\farcs25$ in declination\footnote{The 90\% source 
location error circle in {\it Chandra} has a radius of about 0.5\arcsec, see {\it Chandra} 
Proposers' Observatory Guide v.8.}). 

Next, two PSFs were fitted, now assuming two sources with fixed (offset-corrected)
positions, the correct relative coordinates, and the same PSF width.
Although a very faint extension toward the S is suggested in the {\it Chandra} X-ray image, the best-fit
amplitude at this position is only $(6.5\pm 2.7)$\% of the amplitude of the T Tau N model,
and is therefore at best a marginal detection (at the 2.4$\sigma$ level).

In summary, the {\it Chandra} HRC image (Fig.~\ref{ttauimage}) shows that T Tau N is the
principal contributor to the X-ray source.  This finding is
supported by spectral analysis below that shows low photoelectric absorption compatible
only with T Tau N.
A small contribution from T Tau S cannot, however, be excluded. Because its visual
extinction is $A_{\rm V} > 8$~mag (for T Tau Sb, \citealt{duchene02}), we expect a 
hydrogen column density of $N_{\rm H} > 1.6\times 10^{22}$~cm$^{-2}$, assuming standard 
gas-to-dust ratios \citep{vuong03}.  Some counts could therefore be expected at energies above 
$\approx 1.5-2$~keV only. We estimated an X-ray luminosity of T Tau S
with the PIMMS software\footnote{http://heasarc.gsfc.nasa.gov/Tools/w3pimms.html}, 
adopting $N_{\rm H} =1.6\times 10^{22}$~cm$^{-2}$, an electron temperature of either $\approx 7$~MK or
28~MK (corresponding to either of the higher temperatures found for T Tau N), and 6.1\% of the
total counts. We found $L_{\rm X, 0.3-10~{\rm keV}} = 4\times 10^{30}$~erg~s$^{-1}$ and 
$L_{\rm X, 0.3-10~{\rm keV}} = 1.4\times 10^{30}$~erg~s$^{-1}$, respectively, for the two temperatures.
The true values may be higher if $N_{\rm H}$ is higher, but lower if we overestimated
the count rate contributions in the marginal fit. In any case, given the small contribution compared to T Tau N
and lack of strong significance, we will assume that the observed {\it XMM-Newton} spectra 
are from T Tau N.

\section{The optical, UV, and X-ray light curves}\label{light}

\begin{figure}
\includegraphics[angle=0,width=9.cm]{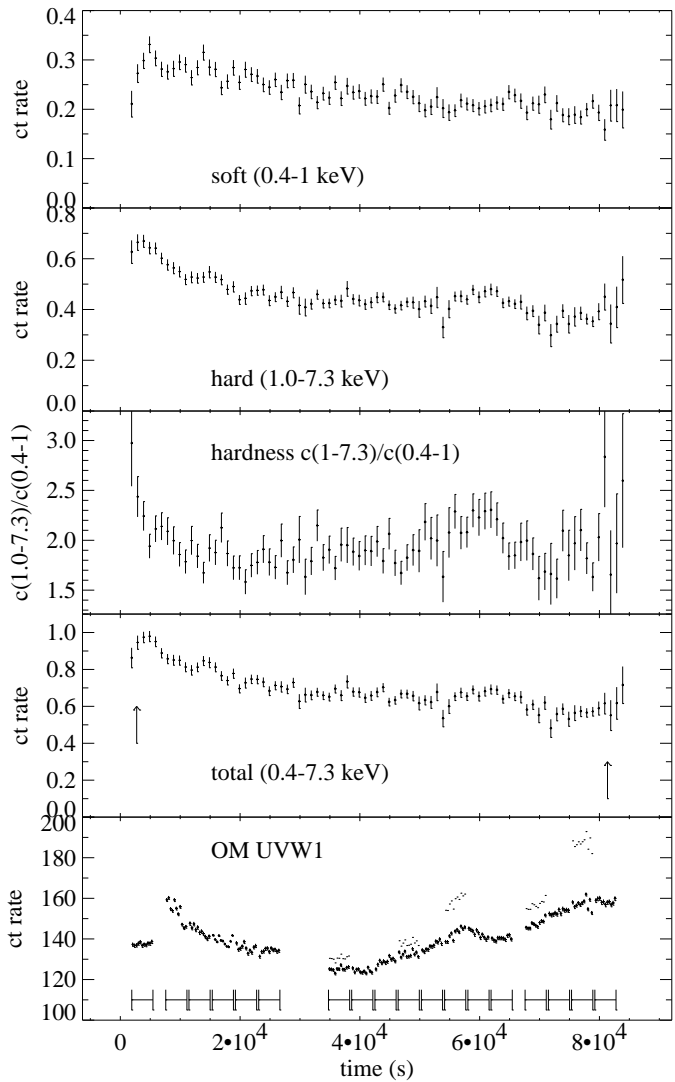}
\caption{Light curves of the T Tau source, extracted from the {\it XMM-Newton} observations. 
{\it From top to bottom:} Combined EPIC PN+MOS1 light curve for soft (0.4--1~keV) counts;
hard (1--7.3~keV) counts; hardness ratio, i.e., ratio between hard and soft count rates;
total light curve (0.4--7.3~keV; the arrows mark the times between which both EPIC cameras were
operational); and the OM UVW1 light curve (dots illustrate bin count rates
before manual correction; see  text for details). Horizontal bars give time
intervals for each OM exposure.}
\label{lightcurve} 
\end{figure}

Fig.~\ref{lightcurve} shows the X-ray and ultraviolet light curves from the {\it XMM-Newton}
observations. 

We extracted the combined EPIC PN and MOS1 background-subtracted light curves for the softer and harder portion of 
the spectrum (energies of 0.4--1.0~keV and 1.0--7.3~keV, respectively; MOS2 was not used because of difficulties
in defining a background curve in the small-window mode).  
These light curves are shown
in the upper two panels of Fig.~\ref{lightcurve}. Both  count rates are slowly decaying, probably from a flare with
a peak that appears to have been recorded at the  beginning of the observation.  The harder
flux decays more rapidly but then also levels off earlier, leading to a hardness increase 
in the second half of the observation, illustrated in the third panel that shows the ratio "hard:soft".
The fourth panel shows the light curve for the entire spectral range. Note that during the first $\approx 1350$~s,
the PN camera did not record data (interval before the time marked by the left  arrow). 
We scaled the MOS1 light curve by an average factor to
the estimated PN+MOS1 count rate. Similarly, only the PN observed during the last 2830~s (time after the second arrow). 
The light curve has again been scaled from the PN count rate to the expected total count rate.

The OM UVW1 light curve is shown in the bottom panel. It was recorded in a series of exposures, each with a duration
of $\approx 3540$~s, followed by a gap of $340$~s. A few exposures were not successfully recorded. 
The high count rate induced large coincidence losses which led to inaccurate calibration
of the incident count rate in some exposures, visible as obvious  discontinuities with respect to the 
count rates in previous or following exposures. The affected exposures were manually corrected by multiplying their count rates
to produce a smoothly connected light curve (Fig.~\ref{lightcurve}; the uncorrected fluxes
are also shown; the count rates in the first exposure were not corrected and are uncertain given the long time gap before
the second exposure). 

Three features are to be mentioned: i) There is no detailed correlation between the X-ray and the UV light curves.
ii) Both curves show, however, a general decay  in the first half of the observation, perhaps as a consequence of
the weak flare seen in X-rays. The time of the UV peak is uncertain, given the unknown correction factor for the
first exposure. It may  have occurred before the start of the observations. iii) The UV light curve increases 
in the second half of the observation, in contrast to the X-rays. However, the X-ray hardness also increases, 
and both UV flux and hardness show a peak around $t = 6\times 10^4$~s. It is possible that both features relate 
to a flare-like heating event that produced both hot coronal plasma (seen in the harder X-ray band) and 
enhanced emission in the chromosphere and transition region (seen in the UVW1 band).

 \begin{figure}
\includegraphics[angle=0,width=9.2cm]{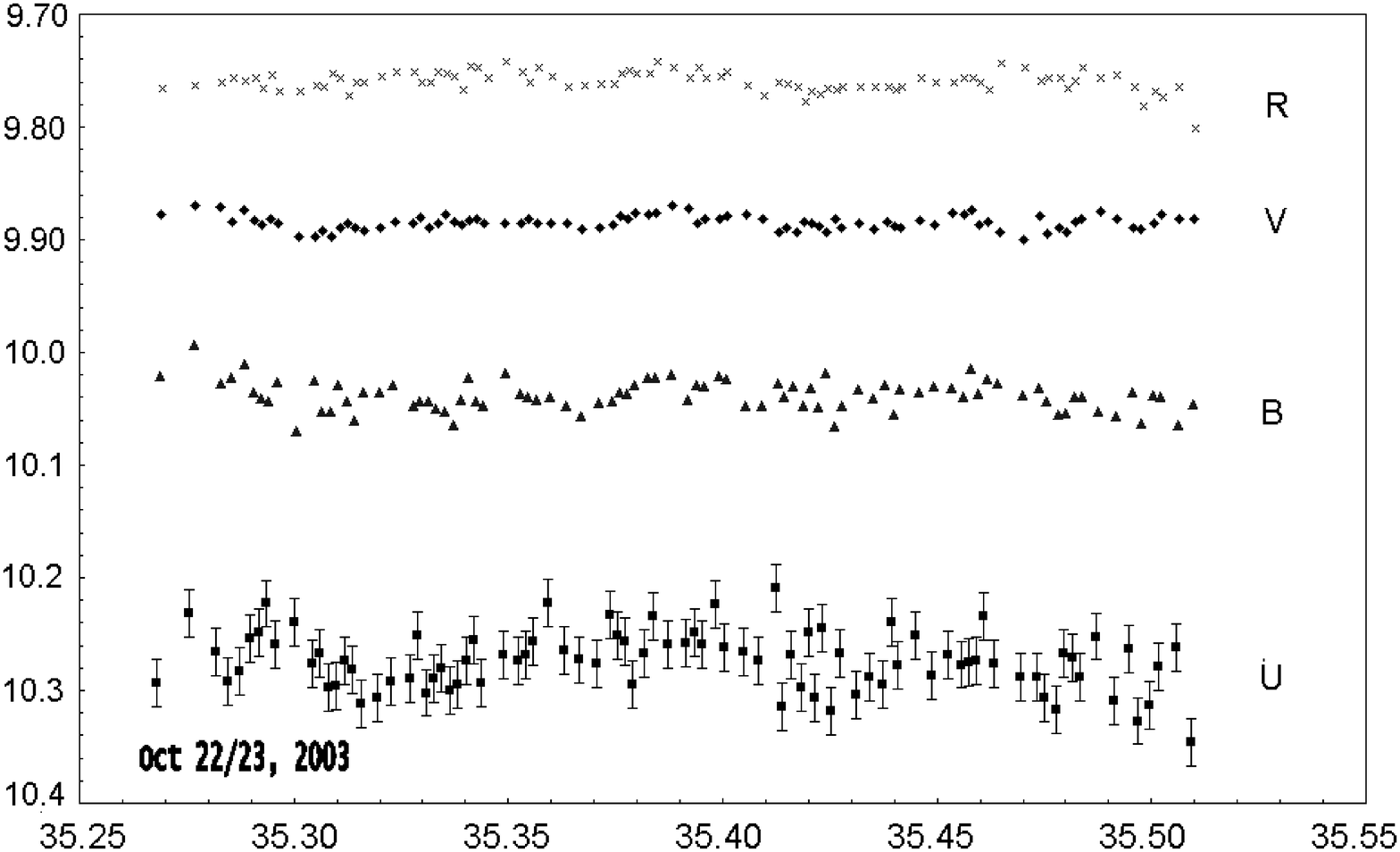}
\includegraphics[angle=0,width=9.2cm]{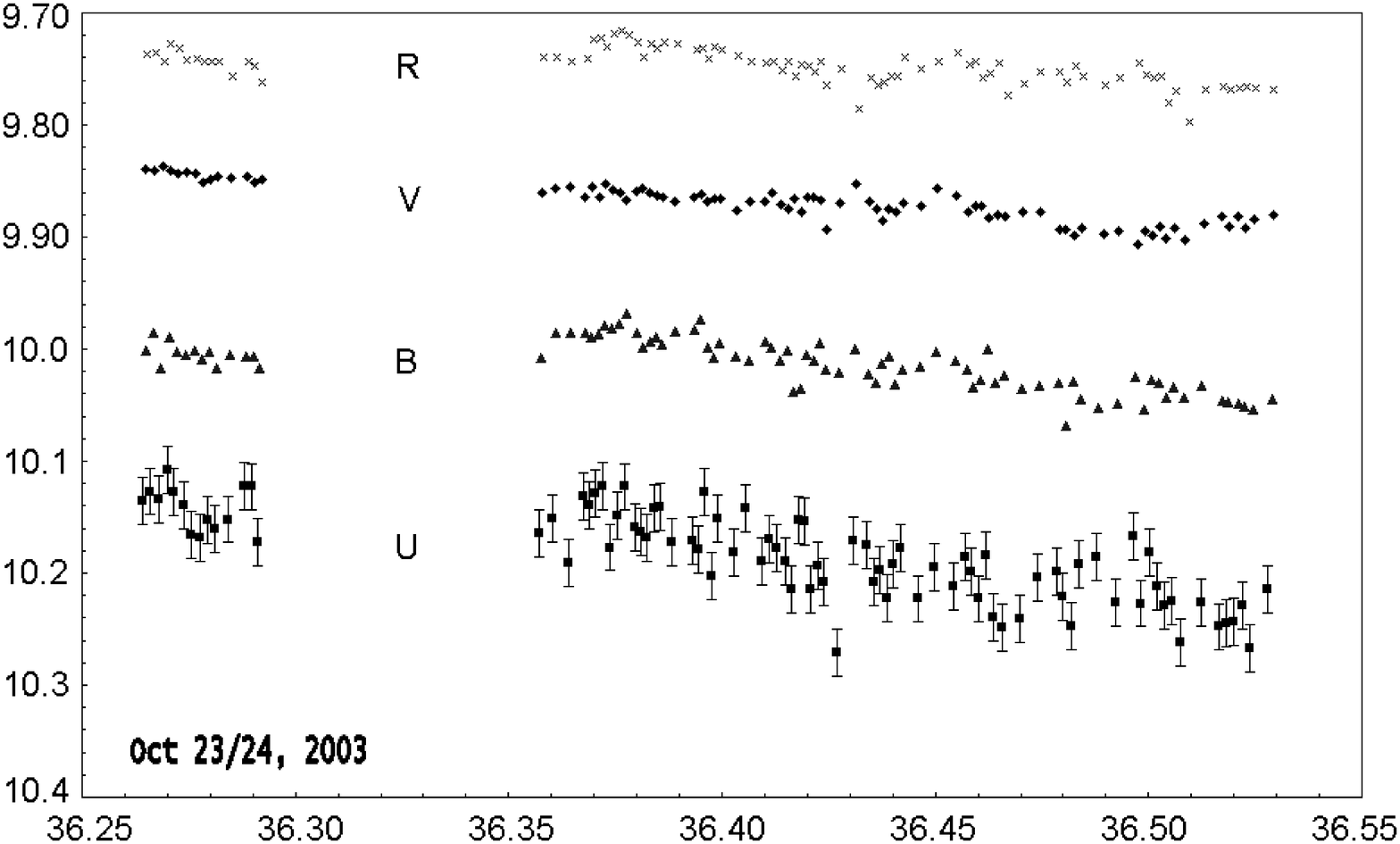}
\includegraphics[angle=0,width=9.2cm]{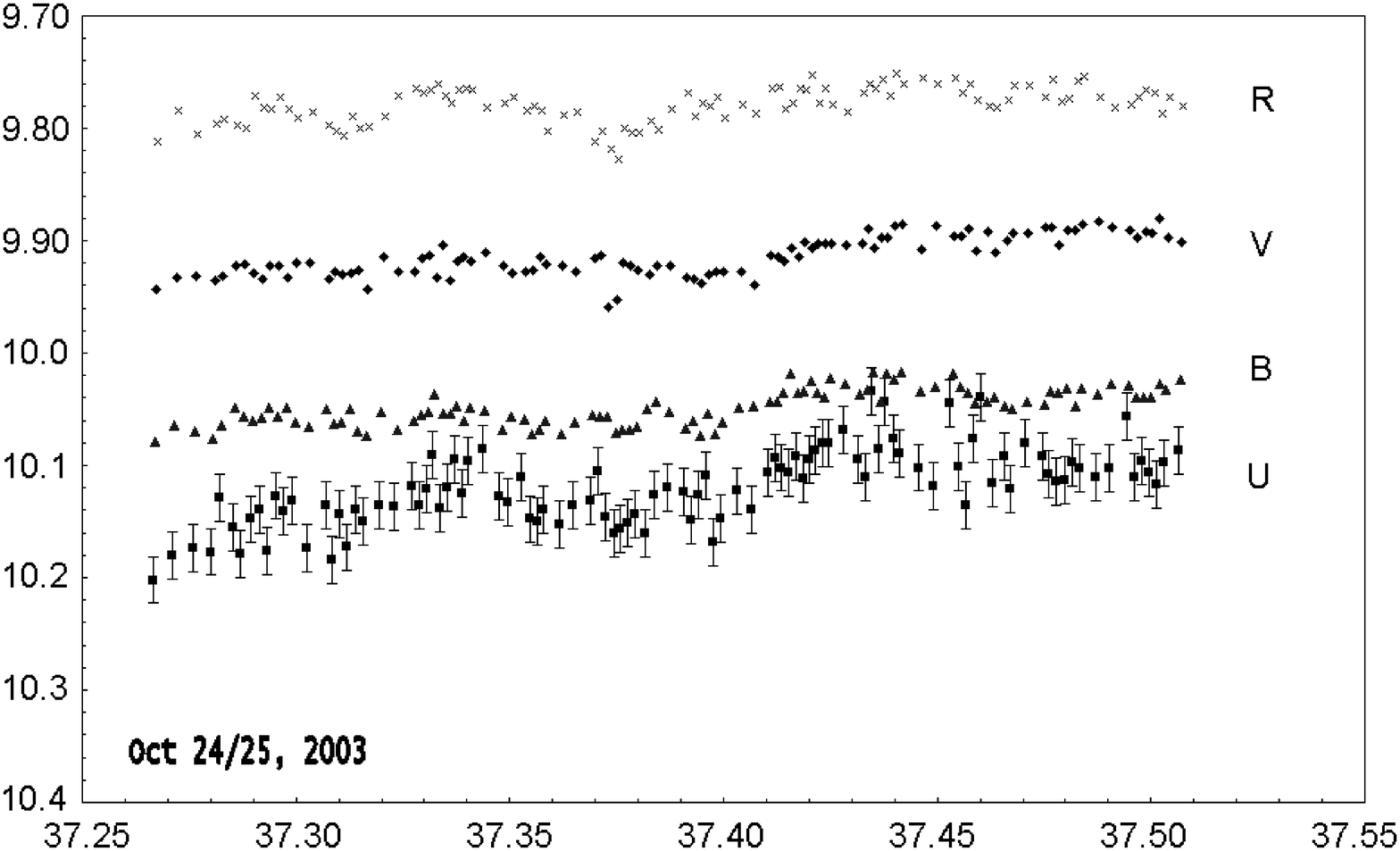}
\caption{Optical UBVR light curves of T Tau, collected during three consecutive nights 
(2003 October 22-25). The abscissae give JD-2452900.0, and the ordinates are in magnitudes. 
Both day-to-day variations and short-term fluctuations (in the U band)
can be seen. }
\label{optical} 
\end{figure}

The light curves in UBVR obtained at Mt. Maidanak Observatory during the ground-based monitoring 
period in 2003 around the
observing campaign with {\it Chandra}  are presented  in Fig.~\ref{optical}. No overlap with the
{\it Chandra} observing interval was achieved. These optical observations  
nearly cover the 2.8-day photometric period of 
T Tau \citep{herbst86} which  corresponds to the rotational period of the 
star. Our  data show day-by-day variability of T Tau: while 
on October 22/23 the brightness of the star is rather stable,  the brightness must have risen in all bands 
between the first and the second observation; prominent slopes   
can be seen in the light curves during the next two days, in opposite directions, first decreasing in brightness,
then increasing again. Given that the rotation period of the star is 2.8~d, at least two minima and two maxima
must have occurred during the stellar rotation covered by these observations. Magnetically active stars
indeed often show two longitude intervals with preferred magnetic activity (e.g., \citealt{berdyugina98}). The 
amplitude of this variability seems to increase with decreasing wavelength 
and therefore reaches its maximum  for the U band. This is particularly evident 
during the October 24/25 observation
when the U-band curve approaches the B-band brightness. This day-by-day variability 
is similar to the long-term variations in the {\it XMM-Newton} UV light curve.
  
\begin{figure}
\includegraphics[angle=0,width=7.5cm]{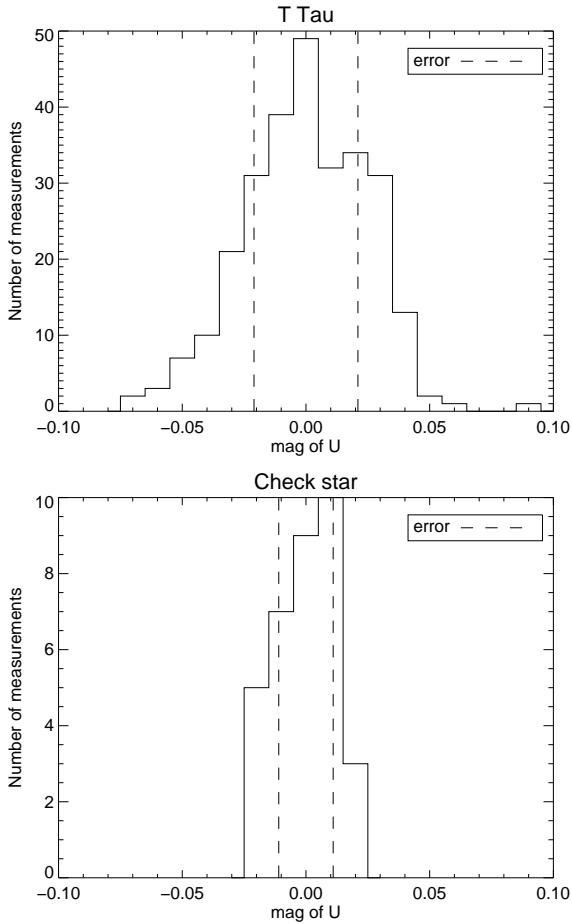}
\caption{Statistical analysis of variability in the U band light curve of T Tau (top) and the
check star HD~284415 (bottom). The solid histograms give the number of measurements with a given magnitude offset from
the average (after removing a slow trend for T Tau). The dashed lines mark the  maximum rms error 
for a measurement at the average magnitude.}
\label{histo} 
\end{figure}

\begin{figure*}
%\centering
\includegraphics[angle=0,width=18.5cm]{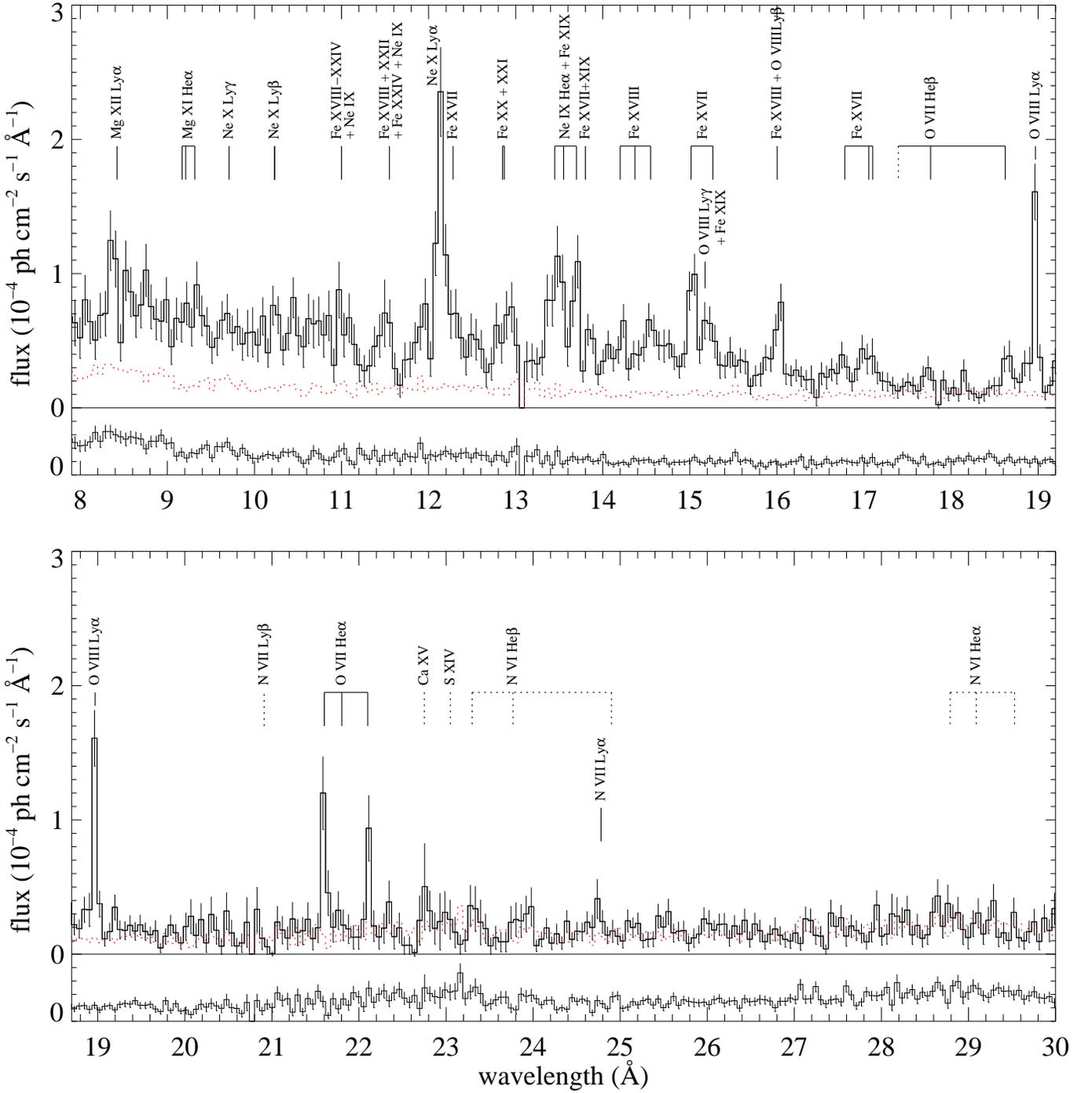}
\caption{Fluxed, combined RGS1+2 spectrum. The background has not been subtracted, but its
spectrum is overplotted as a red, dotted histogram, and also separately shown in the narrow, 
lower panels on the same flux scale.
The bin width is 58~m\AA. The brightest line features contributing to the model fit
are labeled. A few important features that have not been detected are marked with dotted lines.}
\label{rgsfluxed} 
\end{figure*}

Another type of variability which is suggested in the  U-band curve is defined by fast 
small-amplitude variations. In order to estimate the accuracy of our 
observations, we calculated histograms of the brightness distribution for both the check star  
and T Tau itself (Fig.~\ref{histo}). For the second and third T Tau observation, we 
first removed a slow trend in the brightness,
using a polynomial fit that describes the nightly trend but no shorter 
features. The x-axis gives the deviation of points from the mean 
magnitude after removal of the trend. The maximum  rms error of a single measurement 
in the U band calculated from the observations is 0.011~mag for the check star and 0.021~mag 
for T Tau (in the U band the check star was roughly three times brighter than T 
Tau). We plot these errors with dashed lines in the histograms.
The histogram of T Tau suggests that  some of the variability is due to real fluctuations 
beyond statistical errors, while this is not the case for the check star.

\section{High-resolution X-ray spectroscopy}\label{spectroscopy}

We present the combined and fluxed RGS1+2 spectrum in Fig.~\ref{rgsfluxed}.
Given the significant background radiation level, we prefer to first show the full spectrum
before background subtraction and the background spectrum itself separately. The 
potential influence of background features on the spectrum at the source
position can thus be examined in detail. The unsubtracted spectrum also retains maximum
signal-to-noise ratios to reveal weak line features. (We also inspected the spectrum extracted for low-background episodes
only. Apart from a lower S/N, it very closely resembles the spectrum shown here.)
For the subsequent quantitative spectral analysis, we will use background-subtracted RGS spectra 
and provide separate illustrations (see below). We mark all spectral
line features that contained $\ga (3-4)\times 10^{-6}$~photons~cm$^{-2}$~s$^{-1}$, identifying
the dominant ions contributing to the flux (a few of these lines may not be individually 
recognized in the spectrum). Dotted vertical lines point to the location of a few
important fainter lines that remained undetected.

The spectrum  is rich in lines, but also reveals a strong continuum shortward of 17~\AA.
The most unusual aspect of the T Tau X-ray spectrum is its
mixture of features associated with extremely hot plasma, such as the continuum
and several lines of highly-ionized Fe formed above 10~MK,
and lines attributed 
to very cool plasma such as those from N\,{\sc vii} and O\,{\sc vii}.
The ratio between the fluxes in the O\,{\sc vii} He-like line triplet at 21.6-22.1~\AA\
and the O\,{\sc viii} Ly$\alpha$ line at 18.97~\AA\ is near unity. This is unusual
for  magnetically active main-sequence stars and WTTS, while other CTTS have shown similar
ratios, as noted in a survey
of T Tau X-ray grating spectra presented by \citet{telleschi06b}. Among main-sequence solar analogs,
only inactive, old stars reveal such low flux ratios \citep{telleschi05}. We also note
the unusually clear appearance of two He$\beta$ lines of O\,{\sc vii}, at 17.7~\AA\
and at 18.6~\AA\ and the  presence of  the Ly$\alpha$ line of N\,{\sc vii} 
at 24.7~\AA. On the other hand, several features in the unsubtracted spectrum
coincident with the  He$\alpha$ lines of N\,{\sc vi} at 28.5--29.5~\AA\ and the He$\beta$ 
lines of N\,{\sc vi} at 23.3~\AA\ and at 23.8~\AA\ are not significant
because the background spectrum contains similar features at the respective wavelengths.
We emphasize that the identified features of O\,{\sc vii} do not suffer from irregular background.

 We have performed  spectral fits in XSPEC \citep{arnaud96}, using the vapec collisional
 ionization equilibrium thermal line+continuum model combined with a common 
 photoelectric absorption component, parameterized by the neutral hydrogen column density
 $N_{\rm H}$. We use background-subtracted spectra for this analysis. 
 The source spectrum is sufficiently rich in spectral features that 
 a determination of several element abundances is warranted. We note that alternative
 methods such as reconstruction of emission measure distributions from extracted line
 fluxes could in principle be applied; however, most of the important line features
 required for such an analysis are heavily blended, and from no element (including Fe)
 do we see lines from more than two ionization stages at sufficient signal-to-noise ratios.
 We therefore prefer to use template spectra in XSPEC that take into account all
 unresolved blends and faint lines present in the code, while the principal 
 thermal components are determined by the strongest lines and the continuum. A comparison between 
 these two approaches for similar but brighter X-ray spectra from solar analogs has
 shown little difference \citep{telleschi05}.

\begin{figure}
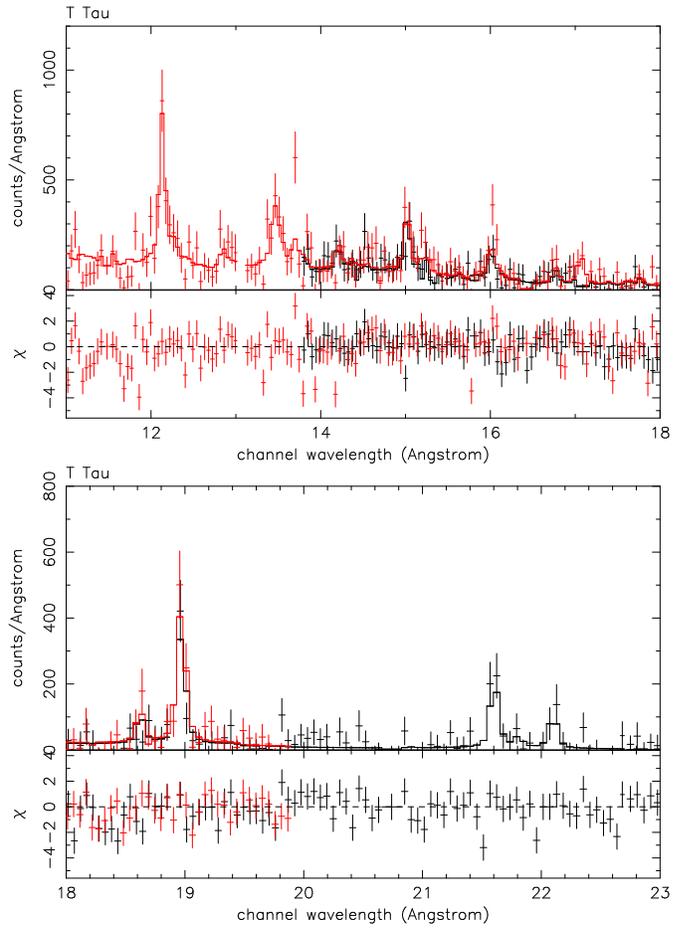

%\centering
\includegraphics[angle=-90,width=8.8cm]{6318fig6a.ps}
\includegraphics[angle=-90,width=8.8cm]{6318fig6b.ps}
\caption{Spectral fit to the background-subtracted RGS1 (black) and RGS2 (red) data in
the 11-18~\AA\ (top) and 18-23~\AA\ (bottom) ranges. Note the missing data in the RGS1 spectrum below 13.8~\AA\
and in the RGS2 spectrum above 20~\AA. These gaps are  due to chip failures earlier in the mission.  The lower panels in each plot show
the residuals in terms of sigmas.}
\label{ttaufit} 
\end{figure}

 We simultaneously fitted spectra of RGS1 and RGS2 between 7~\AA\ and 
 25~\AA, complemented with a portion of the MOS1 spectrum in the range of 1.5--9.35~\AA. 
 We use only one MOS spectrum confined to the harder range because we prefer to
 give full weight to the high-resolution RGS spectra wherever possible. The MOS
 spectrum complements information on the Mg, Si, S, and Fe abundances and constrains the
 hottest plasma components.  All spectra were rebinned so that they contained at least 15 counts per
 bin before background subtraction.
 Three thermal components were required for a good fit, defined by their temperatures $T_{1,2,3}$ and
 the associated emission measures EM$_{1,2,3}$.
 The results are shown in Table~\ref{tab:results}, together with 68\% and 90\% confidence ranges.
 The reduced $\chi^2$ is excellent (1.09 for 382 degrees of freedom = dof). We also list
 the ``average temperature'' $T_{\rm av}$, which has been defined as the logarithmic
  average of the three temperatures, weighted with the respective emission measures; 
  its 68\% range is an approximation calculated using Gaussian error propagation.
  This temperature is considerably lower than $T_{\rm av} = 12.8-17.6$~MK found in the XEST survey 
  from EPIC data \citep{guedel06a}. This is because the EPIC spectral fit does not
  recognize the softest emission for which the cool component at 1.8~MK is responsible.
  If we calculate $T_{\rm av}$ for the two hotter components, we obtain $T_{\rm av} = 15.7$~MK,
  in agreement with the EPIC survey value.
 The fits are illustrated in Fig.~\ref{ttaufit} for two spectral ranges
 of the RGS detectors.
 The RGS data are shown at a higher resolution (fewer counts per bin) than actually used 
 for the fit, in order to better illustrate the individual line features.

\begin{table}
\caption{Results from the 3$T$ model fit }             % title of Table
\label{tab:results}      
\centering            
\begin{tabular}{llll}      
\hline\hline             
Parameter                                           &  best-fit       & 68\% range         &  90\% range  \\    % table heading
\hline                        
$N_{\rm H}$ [$10^{22}$ cm$^{-2}]$                   & 0.49            &  0.46--0.55        & 0.43--0.59 \\
\hline
$T_1$ [MK]                                          & 1.76            &  1.30--2.01        & 0.94--2.16 \\
$T_2$ [MK]                                          & 7.42            &  7.11--7.69        & 6.91--7.87 \\
$T_3$ [MK]                                          & 28.8            &  27.4--30.6        & 26.5--32.3  \\
EM$_1$ [$10^{53}$ cm$^{-3}$]		            & 9.3             &  5.5--17.8         & 3.8--25.8 \\
EM$_2$ [$10^{53}$ cm$^{-3}$]		            & 4.4             &  3.4--6.2          & 2.8--7.7    \\
EM$_3$ [$10^{53}$ cm$^{-3}$]     		    & 5.0             &  4.6--5.4          & 4.1--5.6 \\
\hline
Abundances:$^a$                                     &                 &                    &        	\\
N                                                   & 0.31            &  0.00--2.92        & 0.00--5.62	\\
O                                                   & 0.43            &  0.28--0.75        & 0.28--1.24    \\
Ne                                                  & 0.82            &  0.61-1.02         & 0.50--1.19	 \\
Mg                                                  & 0.49            &  0.34--0.63        & 0.27--0.74 \\
Si                                                  & 0.27            &  0.17--0.37        & 0.12--0.44	 \\
S                                                   & 0.30            &  0.17--0.44        & 0.09--0.54  \\
Fe                                                  & 0.28            &  0.21--0.35        & 0.18--0.40	 \\
\hline
$T_{\rm av}$  [MK]                                  &  5.2            & 4.2--7.3           & ...\\
\hline
$n_{\rm e}$  [$10^{10}$ cm$^{-3}$]                  &  ...            & $<2.6$             & $<8$ \\
\hline
$L_{\rm X, 0.1-10}^b$   [$10^{31}$ erg/s]$^b$       & 2.02	      & ...		   & ...  \\
$L_{\rm X, 0.3-10}^c$   [$10^{31}$ erg/s]$^c$       & 1.52	      & ...		   & ...  \\
\hline
$\chi^2_{\rm red}$                                  & 1.09            & ...                & ...\\
dof                                                 & 382             &...                 & ...\\
\hline                                   %inserts single line
\end{tabular}
\begin{minipage}{0.47\textwidth}
\footnotetext{\hskip -0.5truecm $^a$ Relative to solar photospheric abundances of \citet{anders89} except for
              Fe, which refers to \citet{grevesse99}\\
	      $^b$ for the 0.1--10~keV range, using a distance of 140~pc\\
	      $^c$ for the 0.3--10~keV range, using a distance of 140~pc
	      }
\end{minipage}	      
\end{table}

 The unusual mixture of very hot ($\approx$30~MK) and very cool ($\approx$1.8~MK) plasma
 is confirmed.  In the intrinsic, unabsorbed model spectrum, the O\,{\sc vii} resonance 
 line is in fact {\it the strongest line in the entire X-ray spectrum}, exceeding even the O\,{\sc viii} Ly$\alpha$
 line by a factor of 2.0. Such ratios are otherwise typical only for extremely inactive stars with
 cool coronae, such as Procyon \citep{raassen02}.
 
 The amount of cool emission measure is similar to the total 
 emission measure of the hotter two
 components although the precise ratio strongly depends on $N_{\rm H}$; the latter
 sensitively affects the soft part of the spectrum to which the cool plasma mostly 
 contributes. The value of $N_{\rm H} \approx  4.9^{+0.6}_{-0.3}\times 10^{21}$~cm$^{-2}$ 
 is somewhat higher than found from EPIC data
 alone \citep{guedel06a}, $N_{\rm H} = (3.4\pm 0.1)\times 10^{21}$~cm$^{-2}$ (from a 
 2-$T$ analysis). Part of this discrepancy could be induced by slight, systematic
 calibration inaccuracies at the long-wavelength end of the RGS. Further, for spectra
 of limited S/N, an acceptable spectral fit could  overestimate the cool EM if  $N_{\rm H}$
 is slightly too high.  However, the basic result is evident in the flux ratios of the 
 O\,{\sc vii} and O\,{\sc viii} lines at similar wavelengths. 
 We find a ratio between the  observed fluxes in O\,{\sc vii} and O\,{\sc viii}  of $1.06\pm 0.29$. 
 From the model, we infer that about half of the O\,{\sc viii} Ly$\alpha$ line flux is due
 to the two hotter components. The other half must be due to a cooler component and therefore
 amounts to about 50\% of the total O\,{\sc vii} flux at 21.6--22.1~\AA.   Keeping the hotter 
 two components fixed in the model but enforcing $N_{\rm H} = 3.4\times 10^{21}$~cm$^{-2}$ as in
 the XEST survey (\citealt{guedel06a}, noting that $N_{\rm H}$ of this magnitude principally influences 
 temperature and emission measure of the coolest component), we calculated  conditions for $T_1$ and EM$_1$ 
 to produce the observed flux ratio, given the contributions from the hotter
 components. There is a family of solutions for ($T_1$, EM$_1$)
 below $T_1 = 2.1$~MK, with the smallest emission measure still being 17\% larger than EM$_2$,
 and this solution is obtained for $T_1 = 1.59$~MK. We conclude that even for the lower $N_{\rm H}$,
 the coolest emission measure is comparable to the intermediate and  hotter EMs.

 The element abundances (given with respect to solar photospheric abundances of
 \citealt{anders89} except for Fe for which we used the value of \citealt{grevesse99} as a basis)
 are common among T Tau stars \citep{telleschi06b}. We note
 a relatively large abundance ratio for Ne/Fe (= $2.9\pm1.0$). The abundance of N is very poorly 
 determined, given the very low S/N of any potentially detected line. The usual trend toward an
 inverse first ionization effect in magnetically active stars \citep{brinkman01, guedel01}
 is marginally recovered.
 
 We used the O\,{\sc vii} He-like line triplet at 21.6--22.1~\AA\ to infer electron
 densities in the coolest plasma component. In high-density environments, the forbidden transition 
 ($f, 1s2s\ ^3S_1 \rightarrow 1s^2\ ^1S_0$) at 22.1~\AA\  can become suppressed by electron collisions 
 in favor of the intercombination transitions ($i, 1s2p \ ^3P_{1,2} \rightarrow 
 1s^2\ ^1S_0$) at 21.8~\AA\ (\citealt{gabriel69, blumenthal72}, see \citealt{telleschi06b}
 for applications to high-resolution spectra of other pre-main sequence stars). The 
 density-sensitive range approximately covers $10^{10}~{\rm cm}^{-3} \la n_{\rm e} \la
 10^{12}~{\rm cm}^{-3}$. The limited S/N  makes individual
 extraction of the three line fluxes difficult.  The $i$ line is  not significantly
 detected, while the $r$ and $f$ lines are present (at the 2.3 and 3.6$\sigma$ level, respectively), suggesting low
 densities.  Formally, the  {\it unabsorbed} fluxes at the
 distance of the Earth are $(5.14\pm 1.41, 0.91\pm 0.95, 3.70\pm 1.58)\times 10^{-4}$~photons~cm$^{-2}$~s$^{-1}$
 in the $(r, i, f)$ lines, resulting in a best-fit $f/i$ ratio of 4.0. However, we note that there is
 an additional constraint on the relative line fluxes. The so-called G ratio, $(r+i)/f$, where $r, i,$ and
 $f$ denote the fluxes in the resonance, intercombination, and forbidden lines, is 
 determined by the thermal structure of the plasma. We therefore fixed our best-fit
 thermal model and made use of the density calculations implemented in the vmekal 
 code in XSPEC to infer the densities. We only fitted $n_{\rm e}$ and, for optimum adjustment of
 the line fluxes, the emission measure of the coolest component that predominantly 
 forms the lines. The high $f/i$ ratio requires low densities. The best-fit values are below the density-sensitive 
 range (in agreement with the above $f/i$ estimate) so that we quote only the
 68\% and 90\% upper limits in Table~\ref{tab:results}. These are approximately $2.6\times 10^{10}$~cm$^{-3}$ (68\% confidence)
and $\approx 8\times 10^{10}$~cm$^{-3}$ (90\% confidence) using Gehrels weights on our spectral data.
\footnote{We confirmed these estimates using Monte Carlo simulations
          based on the observed source spectrum, the observed background
          spectrum, the same response matrix and exposure time, and
          the thermal structure and abundances as found in our best
          fit to the entrire spectrum; we obtained an 68\% limit at $2\times 10^{10}$~cm$^{-3}$ 
          and a 90\% limit around $\approx 9\times 10^{10}$~cm$^{-3}$. Conversely,
          if the true density were  $10^{11}$~cm$^{-3}$, we found that only one out
          of ten statistical realizations converges to an inferred
          density below  $\approx 2\times 10^{10}$~cm$^{-3}$.}
 
 The X-ray luminosities, $L_{\rm X}$, in Table~\ref{tab:results} are given for the 
 0.3--10~keV range (adopted for the
 XEST survey, \citealt{guedel06a}) and also for the 0.1--10~keV range to include 
 important contributions from the coolest component. The latter indeed doubles
 $L_{\rm X}$ compared to the EPIC analysis in \citet{guedel06a}. Not considering
 the contributions from the cool plasma, we find $L_{\rm X} = 1.0\times 10^{31}$~erg~s$^{-1}$
 for the 0.3--10~keV range, nearly identical to the XEST survey value of 
 $0.94\times 10^{31}$~erg~s$^{-1}$ based on a two-component spectral fit to the EPIC PN data.

\section{Discussion and conclusions}\label{discussion}

\subsection{Observational signatures}

We now discuss the X-ray properties of T Tau N as found in this paper in the
context of X-ray features that have previously been associated  with 
accreting pre-main sequence stars, and with properties seen in the
U-band and optical light curves. We address, in turn, the following characteristics:
i) Flux ratios in He-like triplets, indicative of electron densities; ii) 
cool plasma;  iii) abundance anomalies; and iv) optical and ultraviolet variability.
We refer to \citet{telleschi06b} and \citet{telleschi06c} for further detailed
discussions of high-resolution X-ray spectra from pre-main sequence stars.

{\bf He-like line flux ratios and electron densities.} High electron densities in cool
plasma could  be a consequence of accretion shocks. The post-shock temperature
in the accretion column near the stellar surface is
$T = 3 v^{2} \mu m_{p}/16k$, where the velocity $v$
is about 0.9 times the free-fall velocity $v_{\rm ff}=(2GM/R)^{1/2}$ \citep{calvet98}, $m_{p}$ is the
proton mass, $k$ is the Boltzmann constant, and the mean molecular weight $\mu
\approx 0.62$ for a fully ionized gas. For T Tau, we use $M = 2.4M_{\odot}$ and $R = 3.6R_{\odot}$ as
listed in \citet{guedel06a}. Then, $v_{\rm ff} = 500$~km~s$^{-1}$ and $T \approx 2.9\times 10^6$~K.
The soft excess is thus compatible with expected post-shock temperatures of the free-falling
gas.

The accretion rate of T Tau is $\dot{M}_{\rm acc} \approx (3-6)
\times 10^{-8} M_{\odot}$~yr$^{-1}$ \citep{white01, calvet04}.
For the accretion  luminosity (neglecting viscous dissipation), we use the expression given
by \citet{telleschi06b}: $L_{\rm acc,30} \approx 600 \tilde{M} \dot{M}_{-8} / \tilde{R}$, where $L_{\rm acc,30}=L_{\rm acc}/
(10^{30}$ erg s$^{-1}$), $\tilde{M}=M/M_{\odot}$, $\dot{M}_{\rm acc,-8}=\dot{M}_{\rm acc}/10^{-8}
M_{\odot}$~yr$^{-1}$  and $\tilde{R}=R/R_{\odot}$. We thus obtain $L_{\rm acc, 30} = 1200 - 2400$, i.e.,  
at least two orders of magnitude more than
required to produce the radiative output in the softest X-rays. Some energy will  
be conducted to cooler layers, and the majority of the X-rays may be entirely absorbed 
\citep{lamzin99}.

Lastly, we can estimate the shock density, using the strong-shock condition $n_2$=4$n_1$, where 
$n_1$ and $n_2$ are the pre-shock and post-shock densities, the former relating to the mass accretion 
rate and the accreting surface fraction $F$ (filling factor - assuming that all flows produce similar shocks). 
\citet{telleschi06b} gave the expression
\begin{equation}
n_2 \approx {4 \times 10^{11}\over \tilde{R}^{3/2}\tilde{M}^{1/2}} {\dot{M}_{-8} \over F}~{\rm [cm^{-3}]}.
\label{eq.density}
\end{equation}
Typical values for  $F$ are $F = 0.1-10$\% \citep{calvet98}. For T Tau,
the density should then be $n_2 = (3.6-7.2)\times 10^9\dot{M}_{-8}/F$~cm$^{-3} = 
(1.1-2.2)\times 10^{11}/F$~cm$^{-3} \ga 1.1\times 10^{12}$~cm$^{-3}$, where the latter value is for the extreme
case of $F = 10$\%, while we obtain  $n_e \approx (1.1-2.2)\times 10^{14}$~cm$^{-3}$ for $F = 0.1$\%. 
\citet{calvet04} derived a filling factor of $F = 3.1$\% specifically for T Tau.  This implies 
$n_e \approx (3.5-7.1)\times 10^{12}$~cm$^{-3}$.
These values are orders of magnitude larger than the densities (or upper limits thereof) 
inferred from the O\,{\sc vii} triplet.  In contrast, the T Tau density upper limit 
($\approx 8\times 10^{10}$~cm$^{-3}$ at the 90\% confidence level) is fully compatible with the measured densities 
of (mostly magnetically active) main-sequence coronae reported by \citet{ness04}. These are distributed between 
$<10^{10}$~cm$^{-3}$ and about
$10^{11}$~cm$^{-3}$ (derived from O\,{\sc vii}), the median and mean $\log n_{\rm e}$ values    
being 10.36 and $10.37\pm 0.06$, respectively (as derived from the Kaplan-Meier estimator analysis
in ASURV which also takes upper limits into account; \citealt{feigelson85, lavalley92}).

We also inspected the Ne\,{\sc ix} density-sensitive triplet formed at somewhat higher temperatures ($\approx$4~MK).
The signal-to-noise ratio is low and the  Ne\,{\sc ix} triplet is strongly blended
with several lines of highly ionized Fe. Nevertheless, there is a clear indication of a
strong $f$ line in that triplet as well, again pointing to low densities (the Ne triplet is density-sensitive in
the range of approximately $n_e \approx 10^{11} - 10^{13}$~cm$^{-3}$).

In summary, then, standard accretion shock physics does not seem to be compatible with a shock-heating
scenario for the soft spectral component forming the bulk of the O\,{\sc vii} (and Ne\,{\sc ix}) lines in T Tau.
We emphasize that T Tau is now the second accreting pre-main sequence star that shows a high $f/i$ flux
ratio, after similar reports on the Herbig star AB Aur \citep{telleschi06c}. This is to be compared with
three CTTS  previously reported to  show high densities, $n_e > 10^{11}$~cm$^{-3}$, namely TW Hya \citep{kastner02},
BP Tau \citep{schmitt05}, and V4046 Sgr \citep{gunther06}. 

{\bf Abundance anomalies.}
High abundances of the volatile elements N and Ne in TW Hya with respect to the Fe abundance have been suggested 
to be due to stronger condensation of Fe and some other elements to dust grains in the circumstellar disk 
\citep{stelzer04}. If the star preferentially accretes from the gas phase, then an elevated Ne/Fe ratio would be measured in 
the coronal or shock-heated plasma.

Anomalously high Ne abundances have been found in the X-ray source of TW Hya \citep{kastner02, stelzer04},
but also in non-accreting evolved stars such as members of the RS CVn class
(Ne/Fe abundance ratios up to 13.4, \citealt{audard03}), and in pre-main sequence stars thought to be
 non-accreting (weak-line) T Tau stars, such as  TWA 5 (Ne/Fe $\approx 10$; \citealt{argiroffi05}) and
HD 98800 (Ne/Fe $\approx 5$; \citealt{kastner04b}). \citet{telleschi06b} have systematically analyzed RGS spectra of 
pre-main sequence stars and compared Fe/Ne and Fe/O abundance ratios for them. They found that i) both ratios 
tend to be higher for young G-type stars than for K- or M-type stars; this trend still holds for near-zero-age main-sequence
stars; and ii) there is no difference between CTTS and WTTS with regard to this trend. 

\begin{figure}
%\centering
\hskip -0.4truecm\includegraphics[angle=-0,width=9.58cm]{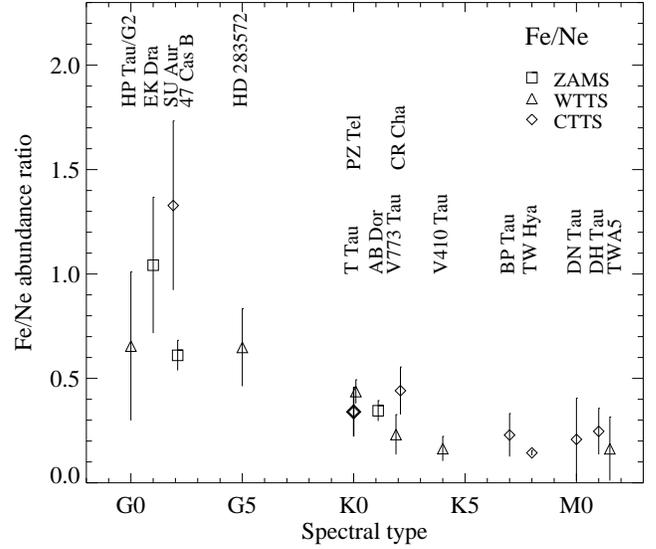}
\caption{The Fe/Ne abundance ratio in the X-ray sources of various pre-main sequence stars
and ZAMS stars as a function of spectral class. The abundance measurements for stars other than
T Tau were taken from the
previous literature and from \citet{telleschi06b} (see that paper for a comprehensive description and references; 
values were converted,
if necessary, to refer to the solar photospheric abundances of \citealt{anders89} and, for Fe, \citealt{grevesse99}). 
The value determined in the present work for T Tau is shown in bold face. Symbols mark different types of stars:
squares: ZAMS stars; triangles: WTTS; diamonds: CTTS.}
\label{nefe} 
\end{figure}

Fig.~\ref{nefe} shows the Fe/Ne ratio derived from high-resolution spectroscopy as a function of spectral type 
for T Tau stars and ZAMS stars (for references, see \citealt{telleschi06b}). The trend for an increasing ratio toward earlier 
spectral types is clearly seen, independently of the evolutionary status (CTTS, WTTS, ZAMS). The ratio for T Tau fits 
in well for a K0 star. The same holds true for the Fe/O ratio (not shown; see \citealt{telleschi06b}). 
We conclude that the abundances observed in T Tau follow a trend that holds for a larger sample
of young stars, and this trend - as judged from a still small sample of stars - does not seem to be
related to accretion properties.

{\bf Soft  excess.} The most outstanding property of the old CTTS TW Hya is its extremely soft X-ray emission \citep{kastner02, stelzer04}.
However, no such X-ray sources have been identified in large samples of CTTS in Orion \citep{preibisch05} or in Taurus 
\citep{guedel06a}. On the other hand, \citet{telleschi06b} have suggested that, apart from a dominant hot plasma, there
is a {\it soft excess} in accreting pre-main sequence stars, best revealed by an unusually large ratio between the (summed) photon fluxes
in the O\,{\sc vii} He-like triplet and the O\,{\sc viii} Ly$\alpha$ line, found to be around  unity for CTTS but typically less than
0.5 for WTTS. The intrinsic ratio between the luminosities of these lines is of course
altered by photoelectric absorption, i.e., the  O\,{\sc vii}/O\,{\sc viii} ratio tends to decrease for increasing
$N_{\rm H}$. The $N_{\rm H}$ value for T Tau compares with the largest in the \citet{telleschi06b} sample, yet the
observed flux ratio is also among the largest. As reported above, we found a ratio of $1.06\pm 0.29$. In the unabsorbed model spectrum,
the O\,{\sc vii} $r$ line is even a factor of two stronger than the O\,{\sc viii} Ly$\alpha$ line, and in fact is the strongest line in the 
X-ray spectrum.

T Tau thus confirms the conjecture that CTTS reveal a soft excess, despite the dominance of an extremely hard component in 
the overall X-ray spectrum. T Tau is  the most extreme case for a soft excess in any T Tau star studied
by high-resolution spectroscopy, except for TW Hya in which the soft component entirely dominates.

{\bf Optical and ultraviolet variability.} Accretion shocks induce excess emission at short optical and ultraviolet wavelengths.
Variability in these bands seen in accreting T Tau stars has therefore been attributed to short-term variations in the mass accretion,
or simply inhomogeneities at certain locations in the streams, rather than to flare-like energy release events  \citep{gullbring96}. 
Variability amplitudes are typically modest (a few tenths of a magnitude) and last for 0.6~hr to several hours. The gas heats
only moderately, again in contrast to chromospheric gas in flares \citep{gullbring96}. The induced variability may be explained
by accretion rate changes of the order of 10\%, which will then require a scale length of the inhomogeneities of $5\times 10^{10}$~cm for
the CTTS BP Tau and, combined with the free fall velocity, correctly explains the variability time scales (see \citealt{gullbring96} for
further details). It is possible that these inhomogeneities reflect in variations of the softest X-ray flux if X-rays are
produced in the accretion shock, but a dedicated monitoring study of BP Tau with ROSAT and ground-based UBVRI photometry revealed
an absence of correlated events; X-ray variability may thus not be related to accretion-induced optical and UV variability \citep{gullbring97}.
New evidence was provided by the comprehensive XEST survey: a comparison of the X-ray light curves of 
many T Tau stars with simultaneous light curves obtained in the U band or a shorter-wavelength ultraviolet band showed little 
correlated behavior, except for a few examples of flares that were observed in both bands \citep{audard06}.

There is mixed evidence in our data. First, both the {\it XMM-Newton} ultraviolet data and the ground-based U-band data show clear evidence
for both day-to-day variations and short-term variability. The former shows time scales of at least half a day, compatible with modulation
of accretion hot spots due to rotation. The short-term events (e.g., at $6\times 10^4$~s in Fig.~\ref{lightcurve}, or at 37.35~d in the bottom panel
of Fig.~\ref{optical}) are similar to those reported by \citet{gullbring96}. Conclusions are ambiguous, however: During the {\it XMM-Newton}
observations, there is a long-term trend in X-ray {\it hardness} that appears to correlate with the UV flux; in particular, there is a decay
from the beginning of the observation that may be related to flare emission.  The harder X-rays seen in our spectra cannot be produced 
by accretion shocks as the infall velocity is too low (see above). Some of the variations of the UV flux may therefore be due
to an excess in chromospheric and transition region heating (producing excess UV emission) in concert with coronal heating (producing
harder X-rays). Also, the short event at  $6\times 10^4$~s  in the UV appears
to be reflected in the X-ray hardness curve. The temporal relation is typical for flares: The ultraviolet excess, a signature of 
impulsive heating of the chromosphere, precedes the X-ray signal, a consequence of filling of magnetic loops with hot plasma. 
Thus, overall a significant contribution to UV variability may come from coronal energy release events, although a clear distinction 
is difficult.  This view is supported by the large survey presented  by \citet{audard06} that shows several obvious X-ray flares with
correlated U-band or ultraviolet features.

Modulation effects due to rotation are also possible. The UBVR observations in Fig.~\ref{optical} show some evidence (e.g., within
the middle or the lower panels, or between the upper and middle panels) for a correlation between B or V magnitude and
the U--B color, indicating that the source is bluer when brighter and redder when fainter. This is expected for both
heated regions and cool spots. Heated regions could be due to either
accretion hot spots or magnetically active chromospheric regions. However, the inclination angle of T Tau is small, $i = 
19-23$~deg \citep{herbst97, eisloeffel98}. Rotationally modulated surface features must therefore be located at latitudes
of approximately $\pm 20$~deg around the equator. The standard magnetic accretion scenario (e.g., \citealt{calvet98})  predicts that accretion
hot spots form at high latitudes. It is therefore more likely that, at least during our UBVR observations, the modulation
was due to active regions (unless the accretion also occurs along more complicated magnetic field structures toward 
equatorial regions). The indication of (at least) two minima and two maxima in Fig.~\ref{optical} could
be due to two preferred active longitudes producing chromospherically active regions near the equator (e.g., \citealt{berdyugina98}). 

\subsection{Implications for models}  

Our discussion thus points to the following conclusions: i) If mass accretion streams induce X-rays in shocks near the surface,
then densities should be much higher than inferred from the observed density-sensitive
line flux ratios. It is possible that some of the observed X-ray emission originates in accretion shocks, but 
only a fraction of the accretion stream, subject to very low densities, would be responsible. ii)  Abundance
anomalies do not seem to reflect accretion properties, and T Tau is no exception. iii) T Tau is, however, the most
outstanding example, apart from TW Hya, of an accreting T Tau star with a soft excess in the X-ray spectrum. Such excess is not found in non-accreting
young stars. We thus suggest that the X-ray production of T Tau is influenced by the accretion process although the X-rays
may not form in the bulk of the accretion footpoints.

Shocks could form in outflows and jets that are related to the accretion process (see \citealt{guedel05} for the
example of DG Tau A). In that case, low temperatures and low densities would be expected. Since T Tau does drive outflows
and is surrounded by Herbig-Haro objects (see references in Sect.~\ref{introduction}), this possibility cannot be excluded,
but explicit evidence is lacking.  The high-resolution {\it Chandra} HRC image (Fig.~\ref{ttauimage}) does not show
any evidence for emission that might be related to a jet, in contrast to what has been found for DG Tau A \citep{guedel05}.
 The entire X-ray spectrum can be explained by a multi-thermal plasma subject to
an overall photoelectric absorption component, again in contrast to the spectra of proposed jet-related X-ray sources
\citep{guedel06b}.  

An alternative location for the interaction between the plasma heating process and accretion is the corona or magnetosphere itself.
Accreting  material is guided along magnetic field lines and can increase the density in these regions considerably (similar to
the estimates presented for the footpoints of the accretion flows, see above). Depending on the stream density and the strength 
of the magnetic fields near the accretion streams, the magnetic geometry may
be considerably shaped by the mass stream itself; magnetic field lines be straightened by the flow, perhaps suppressing 
coronal magnetic energy release so that coronal heating
     may have little or no effect on infalling gas.

If, however, a small fraction of the mass loads onto magnetic fields of active coronal regions, 
it will lower the overall temperature and slightly increase
the density there;  cooling thus also becomes more efficient. A fraction of the coronal
volume affected by a small amount of the infalling material could thus produce a cool coronal component, i.e., an observable soft excess.
A suggestion along those lines was made by \citet{preibisch05} who proposed that the lower $L_{\rm X}/L_{\rm bol}$ ratio of
CTTS compared to WTTS in Orion  is due to accretion flows completely suppressing coronal heating in the respective magnetospheric 
regions.

We estimate what fraction of the accreting material would have to be heated to $T = 1.8$~MK as observed in the softest component, 
assuming that the entire soft component is emitted by such material (rather than by plasma evaporating from the stellar surface).
We provide this estimate for the limiting case that any parcel of inflowing material that becomes subject to coronal heating is heated to $T$ at once, after which
it is permanently cooling by radiation, without further heating. The energy rate to heat accreting, ionized gas to $T = 1.8$~MK is 
$\dot{E} \approx 3k T \dot{N}_{\rm e}$ for hydrogen gas, where $\dot{N}_{\rm e}$ is the rate of new, cool electrons flowing 
into the heating region; there is an equal rate of protons. 
Adopting $\dot{M}_{\rm acc} \approx (3-6) \times 10^{-8} M_{\odot}$~yr$^{-1}$ for the total accretion rate, we find $\dot{N}_{\rm e}   =
(1.1-2.2)\times 10^{42}z$~s$^{-1}$, where $z$ is the fraction of the infalling gas stream that is heated.
Therefore $\dot{E} \approx (8-16)\times 10^{32}z$~erg~s$^{-1}$. We assume that all energy is radiatively 
lost by the plasma at 1.8~MK. Modeling the energy losses of the cool plasma component in XSPEC, we find that only 17\% of the 
radiative power is emitted above 0.3~keV, or 28\% above 0.1~keV. Because the cool component emits $5.2\times 10^{30}$~erg~s$^{-1}$ 
into the 0.3--10~keV band (Sect.~\ref{spectroscopy}), the total radiative energy loss rate of this plasma is  
$3\times 10^{31}$~erg~s$^{-1}$. We conclude that $z = 1.9-3.8$\%  of the accretion flow 
would need to be heated to low coronal temperatures.  This fraction will be smaller if the gas is 
continually heated while radiating.

We conclude, then, that T Tau implies  an important role of accretion in the production of high-energy emission.
We suggest that part of the accreting material adds a cool coronal component by being heated to modest temperatures while flowing into active regions.
We are not able to predict why the X-ray excess is found at temperatures of about 2~MK, but we show
that the additional loading of magnetic fields by cool, accreting mass will suppress heating to the otherwise common coronal temperatures
of $> 10$~MK.

\begin{acknowledgements}
We thank Svetlana Berdyugina for helpful comments on photometry and active longitudes.
We warmly acknowledge financial support by the International Space Science Institute (ISSI) in Bern to
the {\it XMM-Newton} XEST team. X-ray astronomy research at PSI has been supported by the Swiss National Science
Foundation (grants 20-66875.01 and 20-109255/1). 
Part of this research is based on observations obtained with {\it XMM-Newton}, an ESA science mission
with instruments and contributions directly funded by ESA member states and the USA (NASA).
MA acknowledges support by NASA grants NNG05GQ13G and NNG05GF92G for the {\it XMM-Newton} project. In addition, he acknowledges 
support from a Swiss National Science Foundation Professorship (PP002--110504). Support for this 
work was provided by the National Aeronautics and Space Administration through {\it Chandra} 
Award Number SAO GO3-4004X issued by the {\it Chandra} X-Ray Observatory Center, which is operated by the 
Smithsonian Astrophysical Observatory for and on behalf of the National Aeronautics Space Administration 
under contract NAS8-03060.
\end{acknowledgements}


\begin{thebibliography}{}
\bibitem[Anders \& Grevesse(1989)]{anders89}Anders, E.,  \& Grevesse, N. 1989, Geochim. Cosmochim. Acta, 53, 197  
\bibitem[Akeson et al.(1998)]{akeson98}Akeson, R.~L., Koerner, D.~W., \&  Jensen, E.~L.~N. 1998, ApJ, 505, 358
\bibitem[Argiroffi et al.(2005)]{argiroffi05} Argiroffi, C., Maggio, A., Peres, G., Stelzer, B., \& Neuh\"auser, R. 2005, A\&A, 439, 1149
\bibitem[Arnaud(1996)]{arnaud96}Arnaud, K.~A. 1996, in ASP Conf. Ser. 101, Astronomical
               Data Analysis Software and Systems V, ed. G. Jacoby \& J. Barnes
               (San Francisco: ASP), 17
\bibitem[Audard et al.(2005)]{audard05}Audard, M., G\"udel, M., Skinner, S.~L.,
         et al. 2005, ApJ, 635, L81 
\bibitem[Audard et al.(2003)]{audard03}Audard, M., G\"udel, M., Sres, A., Raassen, A.~J.~J., \& Mewe, R. 2003, 398, 1137
\bibitem[Audard et al.(2006)]{audard06} Audard, M.,Briggs, K.~R., Grosso, N., et al. 2006, A\&A, this volume
\bibitem[Berdyugina et al.(1998)]{berdyugina98}Berdyugina, S.~V., \& Tuominen, I. 1998, A\&A, 336, L25
\bibitem[Blumenthal et al.(1972)]{blumenthal72}Blumenthal, G.~R., Drake, G.~W.~F., \& Tucker, W.~H. 1972,
         ApJ, 172, 205
\bibitem[Bouvier et al.(1993)]{bouvier93}Bouvier, J., Cabrit, S., Fern\'andez, M., Mart\'{\i}n, E.~L., \&
         Matthews, J.~M.  1993, A\&A, 272, 176
\bibitem[Brinkman et al.(2001)]{brinkman01} Brinkman, A.~C., Behar, E., G\"udel, M. et al. 2001, A\&A, 365, L162
\bibitem[Calvet \& Gullbring(1998)]{calvet98} Calvet, N., \& Gullbring, E. 1998, ApJ, 509, 802
\bibitem[Calvet et al.(2004)]{calvet04} Calvet, N., Muzerolle, J., Brice\~no, C., et al. 2004, AJ, 128, 1294
\bibitem[den Herder et al.(2001)]{denherder01} den Herder, J.~W., Brinkman, A.~C., Kahn, S.~M., 
          et al. 2001, A\&A, 365, L7 
\bibitem[Drake(2005)]{drake05} Drake, J.~J. 2005, in Cool Stars, Stellar Systems and the Sun 13, ed. F. Favata et al.,
         ESA SP-560, 519
\bibitem[Duch\^ene et al.(2002)]{duchene02}Duch\^ene, G., Ghez, \& A.~M., \& McCabe, C., 2002, ApJ, 568, 771
\bibitem[Duch\^ene et al.(2005)]{duchene05}Duch\^ene, G., Ghez, A.~M., McCabe, C., \& Ceccarelli, C. 
        2005, ApJ, 628, 832
\bibitem[Duch\^ene et al.(2006)]{duchene06}Duch\^ene, G., Beust, H., Adjali, F., Konopacky, Q.~M., \& Ghez, A.~M.  2006, A\&A, 457, L9 
\bibitem[Dyck et al.(1982)]{dyck82}Dyck, H.~M., Simon, T., \& Zuckerman, B. 1982,  ApJ, 255, L103
\bibitem[Eisl\"offel \& Mundt(1998)]{eisloeffel98} Eisl\"offel, J., \& Mundt, R.  1998, AJ, 115, 1554 
\bibitem[Feigelson \& DeCampli(1981)]{feigelson81}Feigelson, E.~D., \& DeCampli, W.~M. 1981, ApJ, 243, L89
\bibitem[Feigelson \& Nelson(1985)]{feigelson85} Feigelson, E.~D., \& Nelson, P.~I. 1985, ApJ, 293, 192
\bibitem[Feigelson \& Montmerle(1999)]{feigelson99}Feigelson, E.~D., \&  Montmerle, T. 1999, ARA\&A, 37, 363 
\bibitem[Feigelson et al.(2006)]{feigelson06}Feigelson, E.~D., Townsley, L., G\"udel, M.,
          \& Stassun, K. 2006, Protostars \& Planets V, eds. B. Reipurth, D. Jewitt, and K. Keil 
	  (Tucson: University of Arizona Press), in press 
\bibitem[Flaccomio et al.(2005)]{flaccomio05}Flaccomio, E., Micela, G., 
         Sciortino, S., et al. 2005, ApJS, 160, 450
\bibitem[Freeman et al.(2001)]{freeman01}Freeman, P.~E., Doe, S., \&  Siemiginowska, A.
          2001, SPIE, 4477, 76
\bibitem[Gabriel \& Jordan(1969)]{gabriel69}Gabriel, A.~H., \&  Jordan, C. 1969, MNRAS, 145, 241
\bibitem[Grevesse \& Sauval(1999)]{grevesse99}Grevesse, N.,  \& Sauval, A.~J. 1999,
        A\&A, 347, 348
\bibitem[Grosso et al.(2006)]{grosso06}Grosso, N., Briggs, K.~R., G\"udel, M., et al. 2006, A\&A, this volume
\bibitem[G\"udel(2002)]{guedel02} G\"udel, M. 2002, ARA\&A, 40, 217 
\bibitem[G\"udel(2004)]{guedel04} G\"udel, M. 2004, A\&AR, 12, 71 
\bibitem[G\"udel et al.(2001)]{guedel01} G\"udel, M., Audard, M., Briggs, K., et al. 2001, A\&A, 365, L336
\bibitem[G\"udel et al.(2005)]{guedel05}G\"udel, M., Skinner, S.~L., Briggs, K.~R.,  et al. 
          2005, ApJ, 626, L53
\bibitem[G\"udel et al.(2006a)]{guedel06a} G\"udel, M., Briggs, K.~R., Arzner, K., et al. 2006a, A\&A, this volume
\bibitem[G\"udel et al.(2006b)]{guedel06b} G\"udel, M., Telleschi, A., Audard, M., et al. 2006b, A\&A, this volume
\bibitem[G\"unther et al.(2006)]{gunther06} G\"unther, H.~M., Liefke, C.,  Schmitt, J.~H.~M.~M., Robrade, J., \&  Ness, J.-U.
         2006, A\&A, 459, L29
\bibitem[Gullbring et al.(1996)]{gullbring96} Gullbring, E., Barwig, H., Chen, P.~S., Gahm, G.~F., \& Bao, M.~X. 1996, A\&A, 307, 791
\bibitem[Gullbring et al.(1997)]{gullbring97} Gullbring, E., Barwig, H., \& Schmitt, J.~H.~M.~M. 1997, A\&A, 324, 155
\bibitem[Herbst et al.(1986)]{herbst86} Herbst, W., Booth, J.~F., Chugainov, P.~F., et al.
         1986, ApJ, 310, L71
\bibitem[ Herbst et al.(1997)]{herbst97} Herbst, T.~M., Robberto, M., \& Beckwith, S.~V.~W. 1997, AJ, 114, 744
\bibitem[Imanishi et al.(2001)]{imanishi01}Imanishi, K., Koyama, K., \& Tsuboi, Y.
         2001, ApJ, 557, 747
\bibitem[Jansen et al.(2001)]{jansen01} Jansen, F., Lumb, D., Altieri, B., et al. 2001, A\&A, 365, L1 
\bibitem[Kastner et al.(2002)]{kastner02}Kastner, J.~H., Huenemoerder, D.~P., Schulz, N.~S.,
         Canizares, C.~R., \&  Weintraub, D.~A.  2002, ApJ, 567, 434
\bibitem[Kastner et al.(2004a)]{kastner04a}Kastner, J.~H., Richmond, M., 
         Grosso, N. et al. 2004a, Nature, 430, 429
\bibitem[Kastner et al.(2004b)]{kastner04b} Kastner, J.~H., Huenemoerder, D.~P., Schulz, N.~S., et al. 2004b, ApJ, 605, L49
\bibitem[Kastner et al.(2006)]{kastner06} Kastner, J.~H., Richmond, M., Grosso, N., et al. 2006, ApJ, 648, L43
\bibitem[Koresko(2000)]{koresko00}Koresko, C.~D. 2000, 531, L147 
\bibitem[Koyama et al.(1994)]{koyama94}Koyama, K., Maeda, Y., Ozaki, M., et al. 1994, PASJ, 46, L125
\bibitem[Lamzin(1999)]{lamzin99}Lamzin, S.~A. 1999, Astron. Lett., 25, 430 
\bibitem[LaValley et al.(1992)]{lavalley92} LaValley, M., Isobe, T., \& Feigelson, E. 1992, in Astronomical
         Data Analysis Software and Systems I, eds. D.~M. Worrall, C. Biemesderfer, \& J. Barnes (San Francisco: ASP), 245
\bibitem[Loinard et al.(2005)]{loinard05}Loinard, L., Mioduszewski, A.~J., Rodr\'{\i}guez, L.~F., et al.
         2005, ApJ, 619, L179
\bibitem[Mason et al.(2001)]{mason01} Mason, K.~O., Breeveld, A., Much, R., et al. 2001, A\&A, 365, L36 
\bibitem[Mel'nikov et al.(2005)]{melnikov05}Mel'nikov, S.~Yu., \& Grankin, K.~N. 2005, Astron. Lett., 31, 109
\bibitem[Momose et al.(1996)]{momose96}Momose, M., Ohashi, N., Kawabe, R., Hayashi, M., \&  Nakano, T.
         1996, ApJ, 470, 1001
\bibitem[Murray et al.(2000)]{murray00} Murray, S.~S., Austin, G.~K., Chappell, J.~H., et al.
          2000, SPIE, 4012, 68 
\bibitem[Ness et al.(2004)]{ness04} Ness, J.-U., G\"udel, M., \& Schmitt, J.~H.~M.~M., Audard, M., \& Telleschi, A.  2004, A\&A, 427, 667
\bibitem[Ness et al.(2005)]{ness05}Ness, J.-U., \& Schmitt, J.~H.~M.~M. 2005, A\&A, 444, L41
\bibitem[Neuh\"auser et al.(1995)]{neuhaeuser95}Neuh\"auser, R., Sterzik, M.~F., Schmitt, J.~H.~M.~M.,
         Wichmann, R., \& Krautter, J. 1995, A\&A, 297, 391
\bibitem[Phillips et al.(1993)]{phillips93}Phillips, R.~B., Lonsdale, C.~J., \&  Feigelson, E.~D. 1993,
         ApJ, 403, L43	 
\bibitem[Preibisch et al.(2005)]{preibisch05} Preibisch, T., Kim, Y.-C., Favata, F., et al. 2005,  ApJS, 160, 401 
\bibitem[Robrade \& Schmitt(2006)]{robrade06} Robrade, J., \& Schmitt, J.~H.~M.~M. 2006, A\&A, 449,737
\bibitem[Raassen et al.(2002)]{raassen02}  Raassen, A.~J.~J., Mewe, R.,  Audard, M., et al. 2002, A\&A, 389, 228
\bibitem[Schmitt et al.(2005)]{schmitt05} Schmitt, J.~H.~M.~M., Robrade, J., Ness, J.-U., Favata, F.,
       \& Stelzer B. 2005, A\&A 432, L35
\bibitem[Skinner \& Brown(1994)]{skinner94}Skinner, S.~L., \& Brown, A. 1994,
         ApJ, 107, 1461	 
\bibitem[Skinner et al.(2003)Skinner, Gagn\'e, \& Belzer]{skinner03}Skinner, S.~L., Gagn\'e, M., \& Belzer,
         E. 2003, ApJ, 598, 375
\bibitem[Smith et al.(2003)]{smith03}Smith, K., Pestalozzi, M., G\"udel, M., Conway, J., \&  Benz, A.~O. 2003,
         A\&A, 406, 957
\bibitem[Solf \& B\"ohm(1999)]{solf99}Solf, J., \& B\"ohm, K.-H. 1999, ApJ, 523, 709
\bibitem[Stelzer \& Schmitt(2004)]{stelzer04}Stelzer, B., \& Schmitt, J.~H.~M.~M.  
         2004, A\&A, 418, 687
\bibitem[Stelzer et al.(2000)]{stelzer00}Stelzer, B., Neuh\"auser, R., \&  Hambaryan, V. 
         2000, A\&A, 356, 949
\bibitem[Str\"uder et al.(2001)]{strueder01} Str\"uder, L., Briel, U.,  Dennerl, K., et al. 2001, A\&A, 365, L18
\bibitem[Telleschi et al.(2005)]{telleschi05} Telleschi, A., G\"udel, M.,  Briggs, K., et al. 2005,
          ApJ, 622, 653 
\bibitem[Telleschi et al.(2006a)]{telleschi06a} Telleschi, A., G\"udel, M., et al. 2006a, A\&A, this volume   % accretion
\bibitem[Telleschi et al.(2006b)]{telleschi06b} Telleschi, A.,G\"udel, M., Briggs, K.~R., Audard, M., \& Scelsi, L.
        2006b, A\&A, this volume                % RGS
\bibitem[Telleschi et al.(2006c)]{telleschi06c} Telleschi, A., G\"udel, M., Briggs, K.~R., et al. 2006c, A\&A, this volume   % AB Aur
\bibitem[Tsujimoto et al.(2002)]{tsujimoto02}Tsujimoto, M., Koyama, K., Tsuboi, Y., Goto, M.,
         \& Kobayashi, N. 2002, ApJ, 566, 974
\bibitem[Turner et al.(2001)]{turner01} Turner, M.~J.~L.,  Abbey, A., Arnaud, M., et al. 2001, A\&A, 365, L27
\bibitem[van Langevelde et al.(1994a)]{vanlangevelde94a}van Langevelde, H.~J., van Dishoeck, E.~F.,
          \&  Blake, G.~A. 1994a, ApJ, 425, L45 
\bibitem[van Langevelde et al.(1994b)]{vanlangevelde94b}van Langevelde, H.~J., van Dishoeck, E.~F.,
         van der Werf, P.~P., \&  Blake, G.~A. 1994b, A\&A, 287, L25 
\bibitem[Vuong et al.(2003)]{vuong03}Vuong, M.~H., Montmerle, T., Grosso, N., et al. 2003, A\&A, 408, 581
\bibitem[Walter(1986)]{walter86}Walter, F.~M. 1986, ApJ, 306, 573
\bibitem[Walter \& Kuhi(1981)]{walter81}Walter, F.~M., \& Kuhi, L.~V. 1981, ApJ, 250, 254
\bibitem[Weisskopf et al.(2002)]{weisskopf02}Weisskopf, M.~C., Brinkman, B., Canizares, C., et al. 2001, PASP, 114, 1 
\bibitem[White \& Ghez(2001)]{white01} White, R.~J., \& Ghez, A.~M. 2001,  ApJ, 556, 265
\bibitem[Wolk et al.(2005)]{wolk05}Wolk, S.~J., Harnden, F.~R. Jr., Flaccomio, E., et al. 2005, ApJS, 160, 423 
\end{thebibliography}
\end{document}